\documentclass[12pt]{article}

\usepackage{latexsym}

\font\tenmsbm=msbm10 scaled 1200
\font\sevenmsbm=msbm9
\newfam\msbmfam
\textfont\msbmfam=\tenmsbm \scriptfont\msbmfam=\sevenmsbm

\makeatletter
\@addtoreset{equation}{section}
\makeatother

\def\be{\begin{equation}}
\def\ee{\end{equation}}
\def\ba{\begin{eqnarray}}
\def\ea{\end{eqnarray}}
\def\bet{\begin{tabular}}
\def\eet{\end{tabular}}

\def\nin{\noindent}

\usepackage[dvips]{graphicx}
\usepackage{amsmath}
\usepackage{amssymb}

\catcode`@=11

\begin{document}

\begin{titlepage}

\vspace{2.5truecm}

\begin{center}

{\Large \bf Gauge approach to the "pseudogap" phenomenology}\\
\vspace{0.5cm}
{\Large \bf of the spectral weight in  high $T_c$ cuprates }\\

\vspace{2.0cm}

P A Marchetti and M Gambaccini\\

 \vspace{2.0cm}
 Dipartimento di Fisica e Astronomia, U. di Padova 
 and INFN, I-35131 Padova, Italy\\

\vspace{2.5cm}

\begin{abstract}
 We assume the $t$-$t'$-$J$ model to describe the $CuO_{2}$ planes of hole-doped cuprates and we adapt the spin-charge gauge approach, previously developed for the $t$-$J$ model, to describe the holes in terms of a spinless fermion carrying the charge (holon) and a neutral boson carrying spin $1/2$ (spinon), coupled by a slave-particle gauge field.
In this framework we consider the effects of a finite density of incoherent holon pairs in the normal state. Below a crossover temperature, identified as the experimental "upper pseudogap", the scattering of the "quanta" of the phase of the holon-pair field against holons reproduces the phenomenology of nodal Fermi arcs coexisting with gap in the antinodal region. We thus obtain a microscopic derivation of the main features of the hole spectra due to pseudogap.
 This result is obtained through a holon Green function which follows naturally from the formalism and analytically interpolates between a Fermi liquid-like and a $d$-wave superconductor behaviour as the coherence length of the holon pair order parameter increases. By inserting the gauge coupling with the spinon we construct explicitly the hole Green function and calculate its spectral weight and the corresponding density of states. So we prove that the formation of holon pairs induces a depletion of states on the hole Fermi surface. We compare our results with ARPES and tunneling experimental data. In our approach the hole preserves a finite Fermi surface until the superconducting transition, where it reduces to four nodes. Therefore we propose that the gap seen in the normal phase of cuprates is  due to the thermal broadening of the SC-like peaks masking the Fermi-liquid peak in the spectral weight. The Fermi arcs then correspond to the region of the Fermi surface where the Fermi-liquid peak is unmasked.

\vspace{0.5cm}

\end{abstract}

\end{center}

\nin PACS numbers: 71.10.Hf, 11.15.-q, 74.72.-h, 74.72.Kf\\
\end{titlepage}

\newpage

\baselineskip 8 mm

\section{Introduction}
The phenomenon of "pseudogap" in hole-doped cuprates appears rather complex, exhibiting different features in various regions of the doping-temperature phase diagram and even the word "pseudogap" is used with different meanings by different authors.
One of the most spectacular features associated with pseudogap phenomenology is the appearance revealed by ARPES of a truncated Fermi surface consisting of Fermi arcs. There have been several theoretical explanations of this phenomenon, many of them critically analyzed in [1], but no consensus has been reached, since none of these proposals explains all the characteristic features of the experimental data.

In this paper we develop an "explanation" of the Fermi arcs within a generalization of a gauge approach to superconductivity in cuprates recently proposed [2], comparing the results with  ARPES and tunneling data.
To understand our proposal it is useful first to sketch the pairing mechanism, eventually leading to superconductivity, for underdoped cuprates presented in [2] within a gauge approach to the $t$-$J$ model: as we dope a vortex-like quantum
distortion of the AF background is generated around the empty
sites (described in terms of fermionic spinless holons) with
opposite chirality for cores on the two N\'eel sublattices.  The spin excitations (bosonic spin-
1/2 spinons) are gapless without doping, corresponding to long-range AF order, but above a critical doping density they acquire a finite gap due to
scattering against the spin vortices and the long-range 
anti-ferromagnetic order is converted to a short-range order.  Due to the no-double occupation constraint, decomposing the hole into holon and spinon generates a local gauge symmetry inducing in turn a gauge attraction between holon and spinon binding them into a
physical hole.

The primary pairing force for the charge carriers is an attraction due to chirality between spin vortices with cores on two different N\'eel sublattices, inducing an attraction between holons in their cores. As a consequence of this attraction
at a crossover temperature, denoted $T_{ph}$, a finite density of
incoherent holon pairs are formed. In [2] it was claimed that this phenomenon produces a reduction of the hole spectral weight on the Fermi surface (FS) and it was proposed to identify this
temperature with the experimentally observed  "upper pseudogap" , where the in-plane resistivity deviates from the linear behaviour. At a lower crossover temperature, denoted $T_{ps}$, also a finite density of incoherent spinon pairs appears, giving rise to a gas of incoherent preformed hole pairs through holon-spinon gauge attraction .

Finally, at a even lower temperature, $T_c$, the hole pairs become coherent and superconductivity emerges.
This approach exhibits yet another crossover temperature [3], $T^*$,
intersecting $T_{ps}$ in the doping-temperature phase diagram.  Such crossover is not directly related to
superconductivity. It corresponds to a change in the holon dispersion.
It is characterized by the emergence of a "small" holon Fermi surface around the
momenta ($\pm \pi/2, \pm \pi/2$), with complete suppression of the (coherent)
spectral weight for holes in the antinodal region and partial
suppression outside of the magnetic Brillouin zone (MBZ).

\begin{figure}[ht]
\centering 
\includegraphics[width=18pc]{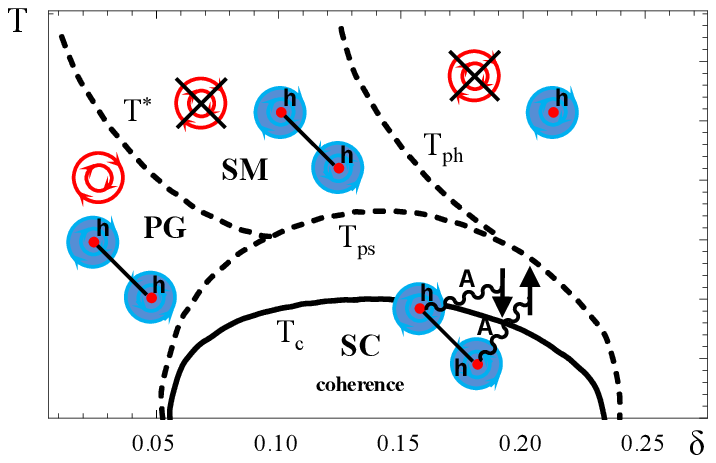}
\caption{Schematic behaviour of the crossovers $T^*, T_{ph}$ and $T_{ps}$ in the $\delta$-$T$ phase diagram. Concentric oriented circles (red online) denote the charge flux, oriented filled disks (blue online) denote the spin vortices, straight lines their pairing attraction, wavy lines the gauge interaction, arrows the spinons.}
\label{fig0}
\end{figure}

  This crossover appears only in bipartite lattices.
Below $T^*$ the effect of
short-range AF fluctuations becomes stronger and the transport physics of
the corresponding normal state region  is dominated by the
interplay between the short-range anti-ferromagnetism of spinons and the thermal
diffusion induced by the gauge fluctuations triggered by the Reizer [4]
momentum. This interplay produces in turn the metal-insulator crossover [3].
We identify $T^*$ in experimental data with the inflection point of
in-plane resistivity and the broad peak in the specific heat coefficient
[5]. The region above $T^*$ in the doping($\delta$)-temperature($T$) phase diagram will be called the ``strange metal phase'' (SM), the one below will be called the ``pseudogap phase'' (PG).
Actually there is no agreement between the experimentalists on the existence of two crossovers (in our approach $T_{ph}$ and $T^*$) associated to the pseudogap phenomenology and the same is true for most of the theoretical approaches.
For clarity in Fig. \ref{fig0} we present a schematic behaviour of the crossovers $T^*, T_{ph}$ and $T_{ps}$ in the phase diagram.

In this paper we extend the pairing mechanism for holes developed for the PG of the $t$-$J$ model to the normal phase of the
$t$-$t'$-$J$ model showing explicitely that $T^*$ and $T_{ph}$ are indeed distinct.

It turns out that a finite density of incoherent {\it holon} pairs induces an angle dependent reduction of the (physical) {\it hole} spectral weight on the FS, starting from the antinodal regions, thus producing a microscopic derivation of the phenomenology of Fermi arcs coexisting with gap in the antinodal region. Notice that this effect on the holes occurs even if the pairing is only among unphysical holons, the spinons being still unpaired, so that there is not yet a gas of preformed hole-pairs.

We show that, when a gas of incoherent holon pairs is present, an energy scale separating low energy modes with a Fermi liquid (FL) behaviour from high energy modes with a $d$-wave superconducting behaviour results naturally and self consistently. This energy scale $m_{\phi}$ will be identified with the the inverse correlation length ("mass") of the quanta of the phase of the holon pairs field.
The scattering of these excitations against holons produces in the holon Green function lowering $T$ a gradual reduction of the spectral weight on the FS at small frequency as we move away from the diagonals of the Brillouin zone. Simultaneously  at larger frequencies we have the formation and increase of two peaks of intensity precursors of the excitations in the superconducting (SC) phase.
The value of $m_{\phi}$ in fact decreases with $T$ , in an extended range approximately linearly, and this decrease drives the system towards the superconducting phase occurring when, possibly discontinously (as experimentally suggested e.g. by [6]), $m_{\phi}$ jumps to 0, in correspondence with the condensation of the hole pairs. Decreasing $m_{\phi}$, the well defined FL quasi-particles appearing at high $T$ gradually lose their coherence in favor of SC-like excitations. The latter gain spectral weight and become increasingly well defined excitations. In the SC phase, all the modes are above $m_{\phi}$ and the holon system is a $d$-wave superconductor, in particular the holon spectral weight at the FS is reduced to zero except on four nodes.

The physical hole is obtained as a holon-spinon resonance produced by the gauge attraction and it inherits the above holon  features, but with a strongly enhanced scattering rate, due to the spinon contribution.

 The behaviour of the spectral weight derived from the above sketched mechanism appears consistent with many ARPES data and it is able to explain the evolution of the density of states as derived in tunneling experiments.
This mechanism of spectral weight suppression exhibits the fingerprint of the presence of the slave-particle gauge field, because the smooth interpolation between FL and SC behaviour is actually due to the interaction of the phase of the holon pairs with the gauge field. Without such interaction the SC-like peaks are strongly suppressed outside of the SC phase, disagreeing with experiments.

Although many phenomenological features in our approach are similar to those of the approach proposed phenomenologically in [7] and partially justified microscopically in [8], at odds with those approaches in ours the hole for $m_{\phi} \neq 0$ (i.e except in the SC phase) has always a FS without gap. This is consistent with the fact that the "pseudogap" region can be reached (at least in our approach) from a FL behaviour without crossing a phase transition, only crossovers being involved in the process, which appears to agree with experiments.
We propose therefore that the gap seen in the normal phase of cuprates is  due to the thermal broadening of the SC-like peaks, the Fermi arcs corresponding  to the region of the Fermi surface where the Fermi-liquid peak is unmasked from that broadening.
A further complete suppression of the spectral weight in the antinodal region occurs in the PG, as previously discussed.
Readers only interested in comparison of final theoretical results  with experiments can skip the next sections and go directly to section 8, where we recapitulate the key points of our approach  before comparing with experimental data.
The effects of the reduction of the spectral weight on transport properties will be discussed in a companion paper [9] where we prove, in particular, a deviation with negative curvature from linearity in $T$ of in-plane resistivity.

\section{ The spin-charge gauge approach to the $t$-$t'$-$J$ model}
The approach in its original version [3] assumed the $t$-$J$ model as model Hamiltonian for the low-energy physics of the CuO${}_2$ planes of the cuprates. Here we add a negative next-nearest-neighbor hopping term $t'$ (with $|t'| < t$) which is known necessary to fit the FS of cuprates. The scheme of the approach, however, remains unchanged and we sketch it here for reader's convenience and to set up the notation, emphasizing only the new features.

 We decompose the hole operator $c_{i \sigma}$ at site $i$ of the  $t-t'-J$ model as $c_{i \sigma} = h_i^* b_{i \sigma}$, where $h$ is a spinless
fermionic holon, carrying charge, while $b_\sigma$ is a spin 1/2
bosonic spinon, carrying spin and obeying the constraint $\sum_{\sigma}
b^{*}_{i\sigma}b_{i\sigma} = 1$. The redundant degree of freedom arising from this decomposition is cured by an emergent
slave-particle $U(1)$ gauge field, $A_\mu$, minimally coupled to holon
and spinon with the same charge. With the choice of statistics adopted  the
no-double occupation constraint is automatically satisfied because of the
Pauli principle for the holon. [ For simplicity in this paper we use the same symbols to denote the field operators in the hamiltonian formalism and the corresponding fields in the path-integral lagrangian formalism. 
] One then uses the possibility offered in 2D (and 1D) to add a ``statistical''
spin flux ($e^{i \Phi^s}$) to $b_\alpha$ and a ``statistical'' charge
flux ($e^{-i \Phi^h}$) to  $h$
``compensating'' each other so that the
product $e^{-i \Phi^h} h e^{i \Phi^s} b$ is still a fermion. The
introduction of these fluxes in the lagrangian formalism is
materialized through  Chern-Simons gauge fields. We then optimize their choice in an improved mean-field approximation (MFA). A key step of MFA is to find a reference spinon configuration with respect to which expand the fluctuations that will be described by a new staggered spinon field, $z_\sigma$. This reference configuration is found
 optimizing the free energy of holons in the presence of a fixed holon-dependent spinon configuration.
 One can show  as in the 1D [10] and 2D [11] $t$-$J$ model that the optimization involves a spin-flip associated to every holon jump between different N\'eel sublattices. Furthermore, if we neglect the spinon fluctuation $z_\sigma$ in the spin flux,  the effect of the optimal spin
flux is to attach a spin-vortex to the holon, with opposite
chirality on the two N\'eel sublattices.
More precisely with this approximation
\begin{eqnarray}
\label{eq:7}
\Phi^s(x) \approx -\sum_l
h^*_l h_l\frac{(-1)^{|l|}}{2}
\arg(\vec{x}-\vec{l}).
\end{eqnarray}
The gradient of $\Phi^s$ can be seen as the potential of a vortex.
These vortices take into account the long-range quantum distortion
of the AF background caused by the insertion of a dopant hole, as first
discussed in [12], the rigidity holding up them being provided precisely by the AF background.

Neglecting the holon fluctuations in $\Phi^h$, the optimal charge flux in the 2D $t$-$J$ model was argued to provide a staggered flux $\pi$ per plaquette for small enough doping and temperature. This anomalous behaviour near half-filling was justified on the basis of a rigorous result by Lieb [13] and a numerical simulation, proving that in a square
 2D lattice at half-filling the optimizing flux for the free energy of fermions is indeed translationally invariant and $\pi$ per plaquette. The contribution from the reference spinon configuration in the optimization can be shown to trivialize the optimal flux at high enough doping or temperature, in agreement with diamagnetic inequality. Although Lieb's proof does not extend to the $t$-$t'$ model, one can still rely on the results of [14] which show that for elementary circuits, triangles and squares, the optimal flux at half-filling is $\pm \pi/2$ for triangles and $\pm \pi$ for squares.  Assuming that the above rule for fluxes matches with the staggered flux $\pi$ per square plaquette of the underlying $t$ model one founds that along the diagonal links of the $t$-$t'$ model the flux should be trivial, this implying that parity and time-reversal are still preserved. On the basis of the above considerations we extend the conjecture made for the $t$-$J$ model as

{\it Assumption} The optimal charge flux in the $t$-$t'$-$J$ model at sufficiently small doping and temperature is given as follows:
let $i$ be a site in the even N\`eel sublattice and $i \pm \hat{\alpha}, \alpha= 1,2$ be its n.n. sites in the 1 and 2 directions respectively and $i \pm \hat{\beta}, \hat{\beta}= \pm \hat{1} \pm \hat{2}$ its n.n.n. sites. Then
\begin{eqnarray}
\label{eq:8}
&&\Phi^h(i \pm \hat{\alpha})-\Phi^h(i)
=(-1)^{\alpha} (\pm i \pi/4) , \quad
\Phi^h(i \pm \hat{\beta})-\Phi^h(i)= 0.
\end{eqnarray}
Also for the  $t$-$t'$-$J$ model the contribution from the reference spinon configuration in the optimization can be shown to trivialize the optimal flux at high enough doping or temperature.
Extending a conjecture made for the $t$-$J$ model we propose that the crossover between the two behaviours discussed above corresponds to the crossover between the ``pseudogap phase''(PG) and the ``strange metal phase''(SM) in the cuprates.
Neglecting the
fluctuations of the gauge field $A_\mu$ which can be reinserted by Peierls minimal substitution, the leading terms of the Hamiltonian can then be written as:
\begin{eqnarray}
\label{tJ} &H \simeq \sum_{n.n.<ij >} (-t) AM_{ij} h^*_i h_j e^{i(\Phi^h_i-\Phi^h_j)}
 +J(1-h^*_ih_i-h^*_jh_j) (1-|AM_{ij}|^2)\nonumber\\
  &+ J h^*_ih^*_jh_jh_i |RVB|_{ij}^2 +\sum_{n.n.n.<<ij>>}t'AM_{ij} h^*_i h_j.
\end{eqnarray}
where $AM_{ij}= Tr (z^*_i e^{i (\Phi^s_i-\Phi^s_j)} z_j)^{(i)}(i)$, with $(i)=1$ (resp.$=*$) if $i$ is in the even (resp. odd) N\`eel sublattice, and it is a kind of
 Affleck-Marston spinon parameter [15] and $RVB_{ij}
 =\sum\epsilon_{\alpha \beta} z_{i \alpha} z_{j \beta}$
 is an RVB spinon singlet order parameter.
The AM/RVB dichotomy in (\ref{tJ}) is due to the spin-flip in the optimization procedure described above 
.
Then we use the following MFA for holons and spinons: in the first term in
(\ref{tJ}) we take $<AM_{ij}> \approx 1$, while in the
second term we replace the hole density by its average and in the
normal state we neglect the third term because of being higher
order in doping ($\delta$).
 A long-wavelength treatment of the second term in
(\ref{tJ}) leads to  a spinon (CP$^{1}$) non-linear sigma model with an
additional term coming from the spin flux,
\begin{equation}
\label{mass} J (1-2 \delta) (\nabla \Phi^s)^2 z^*z,
\end{equation}
where $\partial_\mu \Phi^s (x)=\epsilon_{\mu \nu}\partial_\nu
\sum_{j} (-1)^{|j|} \Delta^{-1} (x - j) h^*_jh_j $ with $ \Delta$
the 2D lattice laplacian. In a quenched treatment of spin
vortices one finds for the average $<(\nabla\Phi^s)^2>=m_s^2 \sim \delta |\log
\delta|$, producing a mass gap for the spinon, consistent with AF
correlation length at small $\delta$ derived from the neutron
experiments [16].

 In the parameter region to be compared with the PG
 of the cuprates  the charge flux  $\pi$ per plaquette
 converts the spinless holons $h$ into
Dirac fermions with small Fermi surfaces centered at $(\pm \pi/2,\pm \pi/2)$: $\epsilon_F \sim
t k_F$ , where $k_F \sim \delta$ is the (average) holon Fermi momentum and $\epsilon_F$ the Fermi energy.
 The holon dispersion is defined in the Magnetic Brillouin
Zone.
In the parameter region to be compared with
the  SM of the cuprates,  where
the optimal charge flux per plaquette is $0$,  we recover a standard tight-binding
``large'' FS ( $\epsilon_F \sim t(1+\delta)$) for the holons, centered at ($\pi,\pi$).
Let us briefly comment how these FS arise. In the $t-t'-J$ model the density of the Gutzwiller projected holes is proportional to the doping $\delta$, the holes corresponding in coordinate space to the empty sites. Substituting the representation of the hole fields in terms of holon and spinon in the lagrangian in the presence of the reference spinon configuration identified by the optimization sketched above, for the holons dressed by the charge flux in the SM one finds the 0-energy level at the position of the FS of the tight-binding unprojected model. This result is compatible with the holon density $\delta$ if 2 holons can occupy the same momenta and the "vacuum" of the model is set at half-filled holon band, namely the holons relevant physically for the projected holes are those corresponding to the deviation from half-filling. Noticing [3] that the (fermionic) holon fields dressed by the charge flux describe semions (with statistics intermediate between fermions and bosons), the first property is guaranteed if we apply to them the Haldane-Wu statistics [17] for semions (as in the 1D $t-J$ model [10]) allowing precisely double-occupation in momentum space for spinless semions. We argue that this MF treatment of holons, although not strictly equivalent to the Gutzwiller projection for holes, nevertheless it is still reasonable for small holon energies, close to the FS .
In the same approximation in the PG one finds for the holons dressed by the charge flux the 0-energy level at the position of the small FS quoted above, the "vacuum" for holons then corresponds to the filling of the lower branch of the Dirac double cones with vertices at $(\pm \pi/2,\pm \pi/2)$ generated by the charge $\pi$-flux background, yielding a result characteristic of a 2D doped Mott insulator.

 Holons and spinons are  coupled by the emergent
gauge field $A$ yielding
overdamped resonances with  strongly $T$-dependent life-time.
This dependence originates from the fluctuations of the transverse
mode of the gauge field, dominated by the contribution of the gapless
holons. Their Fermi surface produces an anomalous skin effect, with
momentum scale
\begin{equation}
\label{reizer}
Q \approx (T k_F^2)^{1/3},
\end{equation}
the Reizer momentum [4].
One can take into account approximately an external frequency $\omega >> T$ replacing $T$ by $\omega$ in the
 life-time and in the Reizer momentum.

Let us now turn to holon pairing.
 Spin vortices centered on holons on the two
N\'eel sublattices have opposite vorticity and this produces an attraction, previously
neglected in the MFA; this is the origin of the attractive force between holons.
Physically it is due to the quantum distortion of the AF background caused
by the holes. We include this effect in MFA by introducing also
the term coming from the average of $z^* z$  in (\ref{mass}). We
get the contribution:
\begin{equation}
\label{zh}
J (1-2 \delta) < z^* z > \sum_{i,j} (-1)^{|i|+|j|} \Delta^{-1}
 (i - j) h^*_ih_i h^*_jh_j,
\end{equation}
where $\Delta$ is the 2D lattice laplacian.
In the static approximation for holons (\ref{zh}) describes  a 2D Coulomb gas
 with coupling constant $\tilde J=J (1-2 \delta) < z^* z >$, with
 $< z^* z > \sim \int d^2q (\vec q^2 +m^2)^{-1} \sim
 (\Lambda^2+m_s^2)^{1/2}-m_s $, where $\Lambda \approx 1$ is a UV cutoff,
and charges $\pm 1$ depending on the N\'eel sublattice. For 2D Coulomb
gases with the above parameters a pairing develops for a
temperature $T_{ph} \approx {\tilde J} /2\pi $, which turns out to
appear in the SM. As it occurs for non-weakly coupled attractive Fermi
systems, below $T_{ph}$ there is a temperature at which the pairs
condense, and it will turn out to be the superconducting
transition temperature $T_c$, see [2]; in-between there is a crossover, $T_{ps}$,
due to spinons, but it will not be discussed here. This omission affects only the lower temperature region for the hole Green function discussed in this paper; in principle the approach is able to derive this correction, but it would highly complicate the calculations without affecting most of the main features. We defer this issue to a future publication.

\section{Low energy Hamiltonian for holons}
In this section we discuss in detail the pairing among holons in the SM region of the $t-t'-J$ model in the BCS approximation, adapting the framework developed in [2], in turn inspired by [18]. A brief comment on the modifications needed for the PG region are added at the end. In the next section we incorporate in this framework the fluctuations of the phase of the holon-pairs order parameter.

We start treating the kinetic hamiltonian
\begin{eqnarray}
\label{holontprime}
H_0^h = -t \sum_{\langle ij\rangle}(h^*_i h_j + h.c.)
-t' \sum_{\langle \langle ij\rangle \rangle}(h^*_i h_j + h.c.)-\mu \sum_{i} h^*_i h_i.
\end{eqnarray}

in the two sublattices (even $A$ and odd $B$) scheme, defining $h_{i}=a_{i}$ and $h_{i+\hat{1}}=b_{i}$ for $i\in A$,
as this will be useful in the later treatment in presence of holon pairing that distinguishes the two N\'eel sublattices.

 The two fields defined within the magnetic Brillouin zone (MBZ) and diagonalizing $H_0^h$ are

\begin{eqnarray}
\label{fipiumeno}
\psi_{\pm}(\vec{k})=a_{\vec{k}}\pm e^{ik_{1}}b_{\vec{k}}
\end{eqnarray}

where $\vec{k}\in$ MBZ and the Fourier transforms $a_{\vec{k}}$ and $b_{\vec{k}}$ are periodic outside the MBZ. The energy eigenvalues are $\epsilon_{\pm}(\vec{k})=-\mu+t'_{\vec{k}}\pm t_{\vec{k}}$, where $t_{\vec{k}}=-2t[\cos(k_x)+\cos(k_y)]$ and $t'_{\vec{k}}=-4t'\cos(k_x)\cos(k_y)$.

\begin{figure}[ht]
\centering
\includegraphics[width=12pc]{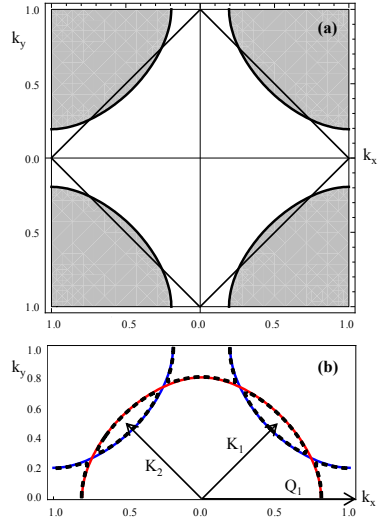}
\caption{Panel (a) represents the FS of the $t$-$t'$ model (t=0.31eV, $|t'|$=0.08eV, $\mu$=0.15eV) defined in the whole BZ (coinciding with that of the $t$-$t'$-$J$ model in the  MFA discussed in the text in the SM); the holes fill the shaded region. The boundaries of the MBZ are also shown. Panel (b) represents the upper rectangular part of the BZ  equivalent to the MBZ after the translation of the 3rd and 4th quadrants with the original FS (blue lines online) and the translated one (red lines online); dashed lines are qualitatively the closed FS of $\Psi_{+}$ (centered in $\vec{K}_{1}$ and $\vec{K}_{2}$) and $\Psi_{-}$ (centered in $\vec{Q}_{1}$)in the presence of holon-pairing. Axis are measured in units of $\pi$.}
\label{fig1FS}
\end{figure}

Both fields have pieces of FS within the MBZ (see Fig. \ref{fig1FS}, panel (a) ). $\psi_{+}$ has four hole-like (FS increases as doping increases)  arcs centered at $\vec{K}_{i}$, in the middle of the sides of the boundary of the MBZ, while $\psi_{-}$ has  electron-like (FS reduces as doping increases) Fermi arcs near $\vec{Q}_{i}, i=1,...,4$, at its vertices. 
In terms of the fields defined in the MBZ the holon field, defined in the whole BZ, turns out to be

\begin{eqnarray}
\label{holpsif}
h_{\vec{k}} =  \left\{
  \begin{array}{ll}
    \psi_{+}(\vec{k}) ,\hspace{0.2cm} \textrm{if}\hspace{0.2cm} \vec{k} \in {\rm MBZ},\\
    \psi_{-}(\vec{k}-2\vec{K}_{v}) ,\hspace{0.2cm} \textrm{if}\hspace{0.2cm}  \vec{k} \not\in {\rm MBZ}, \vec{k} \in BZ.
  \end{array}
\right.
\end{eqnarray}

with $\vec{K}_{v}$ ($v=1,2,3,4$) chosen to keep the argument of $\psi_{-}$ inside the MBZ (where it is defined).

Now we exploit the two reciprocal primitive vectors $\vec{Q}_{\pm}=(\pm \pi,\pi)$ to translate the 3rd and 4th quadrants respectively in order to obtain a upper rectangular zone $D=\{k_x \in [-\pi,\pi]$, $k_y \in [0,\pi]\}$ equivalent to the MBZ, as done in the PG (Fig. \ref{fig1FS} panel (b)).  Notice that each translation exchanges the $\pm$ index of the fields, because of the minus sign due to the exponential factor in eq. (\ref{fipiumeno}), and of the eigenvalues because $t_{\vec{k}+\vec{Q}_{\pm}}=-t_{\vec{k}}$.

In the presence of holon pairing the field $\psi_{+}$ extended by periodicity, restricted to the right quadrant, $D_R$,  and  to the the left quadrant, $D_L$, in $D$ is denoted by $\Psi_{+,\alpha}, \alpha=R,L$. It has a good continuum limit because it has a closed hole-like Fermi surface centered at  $\vec{K}_{1}$ in $D_R$ and at  $\vec{K}_{2}$ in $D_L$, as discussed in a slightly different framework in [19].

Even if mostly neglected in what follows, the field $\psi_- $ extended by periodicity, and then denoted by $\Psi_- $, has a small closed electron-like Fermi surface around the point $\vec{Q}_{1}$, see Fig. \ref{fig1FS} panel (b). It is however convenient to restrict also  $\Psi_- $ to  $D_R$ and $D_L$ to treat it consistently with  $\Psi_{+,\alpha}$, these restrictions are denoted by  $\Psi_{-,\alpha}, \alpha=R,L$.
We now discuss the holon attraction arising from (\ref{zh}). This interaction is able to distinguish between the two sublattices, therefore, when incoherent holon pairs appear the description in the MBZ is appropriate. However the pieces of the original FS of the Nearly Free Electron description will deform to reach orthogonally the boundary of the MBZ as a consequence of the interaction between the previously defined segments of Fermi surface inside and outside the MBZ, and the hole-like and electron-like FS discussed above split-off, the fields $\Psi$ having a closed FS. In the BCS treatment adopted in the following the deformation of the FS is concentrated near the boundary of the MBZ, where hole-like and electron-like FS are closer.
Finally we point out the relevance in our discussion of the negative next-nearest-neighbor hopping term $t'$ which bends the FS of holons and allows BCS pairing of quasi-particles having a good continuum limit in the SM region. In the simple $t$-$J$ model (where $t'=0$), the FS in the SM does not cross the boundaries of the MBZ and our approach to the phase fluctuations of the holon pair order parameter presumably is possible only in the PG region where closed Fermi surfaces centered at $\frac{1}{2}(\pm \pi,\pm \pi)$ arise.
Therefore we guess that in the pure $t$-$J$ model the formation of holon pairs only appears close to $T^*$ and not at the naively deduced $T_{ph}$, which instead can give a rough estimate of the crossover in the full $t$-$t'$-$J$ model.

The structure of FS for holons discussed above will be qualitatively inherited by physical holes, via gauge coupling to spinons discussed later.  Such structure bares some resemblance with that appearing in the spin-density wave approaches [19,20], but in our case its origin is the holon-holon pairing interaction distinguishing the two sublattices, not directly the standard AF interaction, although since the holon pairing originates from the $J$ term it is still of AF origin. Coexisting hole- and electron-like FS at intermediate dopings appear also in large-U treatments of the 2D Hubbard model in terms of an effective $t$-$t'$-$t"$-$J$ model [21]. In both cases going from low to high doping level, one first find small pockets like in our PG, then coexistence, like in our region between $T^*$ and $T_{ph}$ and finally a large FS, as above $T_{ph}$ in the SM.

The attractive interaction between holons in different sublattices, hence with opposite vortex chirality, of eq. (\ref{zh}) is treated at large scales in analogy with the treatment in the PG for the $t$-$J$ model. Since not all vortices form pairs, a finite screening effect persists and the gas of vortices still have a finite correlation length [22], which we denote by $\xi \approx (J k_F)^{-1/2}$, where $k_F$ is the average Fermi momenta of the corresponding closed FS. We keep track of the screening effect by replacing in the long wavelength limit $\Delta^{-1}$ in eq. (\ref{zh}) by an effective potential $ V_{\text{eff}}(\vec{q}) \approx g/ (q^2 + \xi^{-2})$. Hence we approximate the interaction hamiltonian at large scales as:

\begin{eqnarray}
\label{smintholons}
H^h_{I}  =   - \sum_{i,j} V(\vec{i}-\vec{j})
a^{*}_i b^{*}_j b_j a_i
 \approx   - \tilde{J}
\sum_{\vec{p},\vec{q}} V_{\text{eff}}(\vec{p}-\vec{q})
a^{*}_{\vec{p}} b^{*}_{-\vec{p}}
b_{-\vec{q}} a_{\vec{q}}
\end{eqnarray}

We perform the translation discussed above, add right $R$ and left $L$ labels ($\alpha=R,L\equiv 1,2$) to distinguish the two hole-like FS in $D$ and we measure momenta from $\vec{K}_{R}\equiv \vec{K}_{1}$ and $\vec{K}_{L}\equiv \vec{K}_{2}$ respectively. We neglect the interaction between $R$ and $L$ sectors, except for the effect of deformation near the MBZ boundary discussed above and taken into account only phenomenologically, and we adopt the BCS approximation for the $R$ and $L$ sectors defining the order parameter

\begin{eqnarray}
\label{horderp}
\Delta^h_{\alpha,\vec{k}} = \tilde{J} 
\sum_{\vec{q}} V_{eff}(\vec{k}-\vec{q})
\langle b_{\alpha,-\vec{q}} a_{\alpha,\vec{q}}\rangle.
\end{eqnarray}

For the fields $(\Psi_+,\Psi_-,\Psi_-^*,\Psi_+^*)$ the Hamiltonian kernel appears in block-diagonal form:
\begin{eqnarray}
\label{matrb}
\left(
    \begin{array}{cccc}
     -\mu -t'_{\alpha,\vec{k}} -t_{\alpha,\vec{k}} & 0 & 0 & -\Delta^{h}_{\alpha,\vec{k}} \\
    0 & - \mu -t'_{\alpha,\vec{k}} +t_{\alpha,\vec{k}} & \Delta^{h}_{\alpha,\vec{k}}  & 0 \\
    0 &\Delta^{h*}_{\alpha,\vec{k}} & \mu +t'_{\alpha,\vec{k}} -t_{\alpha,\vec{k}} &0 \\
    -\Delta^{h*}_{\alpha,\vec{k}} &0& 0&\mu +t'_{\alpha,\vec{k}} +t_{\alpha,\vec{k}}
    \end{array}
  \right)
\end{eqnarray}
with

\begin{eqnarray}
\label{tetprimo}
t_{\alpha,\vec{k}} = t_{\vec{k}+\vec{Q_{\alpha}}} = -2t [\sin k_x -(-1)^{\alpha}\sin k_y], \quad
t'_{\alpha,\vec{k}} = t'_{\vec{k}+\vec{Q_{\alpha}}} = -4t'(-1)^{\alpha} \sin k_x \sin k_y
\end{eqnarray}
provided  $\Delta^{h}_{\alpha,\vec{k}}$ is odd in $\vec{k}$; the symbols $\pm$ are omitted here and often in the following, when their
presence can be obviously understood. The choice of the order parameter with lowest angular momentum and hence energy has then $p$-wave symmetry. We assume that this order parameter has $p_x + p_y$-wave symmetry for $\Psi_{+,L}$ and $p_x - p_y$-wave symmetry for $\Psi_{+,R}$, so that
\begin{eqnarray}
 \Delta^{h}_{\alpha,\vec{k}}=\Delta^{h}_\alpha({|\vec{k}|}) \gamma_\alpha({\vec{k}}), \quad
 \gamma_\alpha({\vec{k}})=\sin k_x +(-1)^{\alpha}\sin k_y
\end{eqnarray}
with $\Delta^{h}_{\alpha}(|\vec{k}|)$ assumed constant, and denoted  $\Delta^{h}_{\alpha}$, near the FS in the BCS approximation. From eq.(\ref{horderp}) we get $\Delta^{h}_{+,R}=\Delta^{h}_{+,L} \equiv \Delta^{h}_{+}.$
Then we obtain for the field $h$ a $d$-wave symmetry gluing $R$ and $L$ sectors when the full BZ is restored (as we see from its definition, the order parameter changes its sign when the 3rd and 4th quadrants are translated). This $p$-wave symmetry in $D$ generates also an $s$-wave pairing for the electron-like Fermi surface of $\Psi_-$, as discussed in a similar situation in [19].

The energy spectrum of the quasi-particles described by $\Psi_{+,\alpha}$ has the BCS form $E_{\alpha,\vec{k}} = \pm \epsilon_{\alpha,\pm, \vec{k}}$ where

\begin{eqnarray}
\label{smholpectra}
\epsilon_{\alpha,\pm, \vec{k}} = \sqrt{(\pm t_{\alpha,\vec{k}}-t'_{\alpha,\vec{k}}-\mu)^2 + |\Delta_{\alpha,\vec{k}}|^2}.
\end{eqnarray}

Let us briely comment on the PG, where
\begin{eqnarray}
t_{\alpha,\vec{k}} = -2 t \sqrt{(\sin k_x )^2+ (\sin k_y)^2}.
\end{eqnarray}
measuring the momenta from $\vec{K}_{R}$ ($\vec{K}_{L}$) in the $R$ ($L$) sectors of $D$. 
In the PG the analog of the $\Psi_-$-field in $D$ has no FS, but there is a strong matrix effect due to the Dirac structure of the holon field, eventually leading for the hole to an angle-dependent wave-function renormalization constant, $Z(k_x, k_y) \approx 1-(\sin k_x-(-1)^\alpha \sin k_y)/(\sqrt {2} ((\sin k_x )^2+ (\sin k_y)^2)^{1/2})$  in $D_\alpha$, even in absence of holon pairing [3].
 For the holon pairing the main differences between the SM and the PG  are the shape of the FS and the higher value of the (average) Fermi momentum of the former w.r.t. the latter ($k_F \sim \delta$), with an induced difference for the modulus of the holon-pair field. The Fermi velocity instead is roughly the same.

\section{Integration of high energy modes}

In this section we turn to the path-integral formalism and derive the low energy effective action which describes the pairing process in the normal state as we lower the temperature until the superconducting transition is reached.

We start from the spinon sector. The relevant action is the non-linear $\sigma$-model action with a mass term discussed in section 2. This allows us to safely integrate out spinon degrees of freedom obtaining an effective action for the slave particle gauge field $A$. By gauge invariance the leading order is a Maxwell-like action:

\begin{eqnarray}
\label{gaugefspinon}
S_{\text{eff}}^{s}(A_{\mu})=\frac{1}{3 \pi m_s}
\int_{[0, \beta] \times {\bf R}^2} d^3x F_{\mu\nu}^2.
\end{eqnarray}

For the holon sector, in the BCS approximation discussed in section 3, the holon is
gapless only at the 4 nodal points of $\Delta^h_{\vec{k}}$. However in a
large-scale gauge-invariant treatment whereas one can keep constant the modulus
of the order parameter $\Delta^h$ near the FS as in BCS, we must include its
spatially dependent phase, which we denote by $\phi^h_\alpha(x)$ . This is done setting:
\begin{equation}
\label{orpaxsp}
\Delta^h_{\alpha}(\vec{x},x_0)=\Delta^h e^{i\phi^h_\alpha(\vec{x},x_0)},
\end{equation}
(A precise
procedure to go from the lattice to the continuum phase field is
discussed in [23].) The effects of $\phi^h(x)$ on holons is
non-trivial, as first suggested in a different setting in [24], and will be discussed here in detail.
We generalize the BCS interaction between the holon pairs of section 3 and the FL quasi-particles by allowing phase fluctuations of $\Delta^h$.
The holon action for $\Psi_+$ near the FS in terms of Matsubara frequencies $k_0$ then reads:

\begin{eqnarray}
\label{holactintcond}
&&S^{h,\Delta} (\Psi_{\alpha},\Delta^h_{\alpha}) = \sum_{\alpha,k}[ik_0 - E(\vec{k})+ \mu]\Psi^{*}_{\alpha,k}
\Psi_{\alpha,k}
\\
&& -\sum_{\alpha,k,p} \frac{1}{2} \left\{ \Delta^h_{\alpha}(k+p)[\gamma_{\alpha}(\vec{k})-\gamma_{\alpha}(\vec{p})] \Psi^{*}_{\alpha,k}\Psi^{*}_{\alpha,p} + c.c.\right\}\nonumber
\end{eqnarray}

where we put $k=(k_0,\vec{k})$. Linearizing the dispersion we set  $E(\vec{k})-\mu \approx v_F(|\vec{k}|-k_F(\theta))$ near the FS, with $\theta$ an angle parametrizing the FS. We notice that in the case of constant order parameter $\Delta^h_{\alpha}(\vec{k})\rightarrow \Delta^h \delta^2(\vec{k})$, Eq. (\ref{holactintcond}) reproduces the standard BCS coupling discussed in section 3.

One could then reinsert in the kinetic term the gauge field by Peierls subsitution  and naively one would integrate out the holons, again considering the leading term in $A$. However this is not correct since, as will be shown in section 5, above $T_c$ the phase fluctuations have short-range decay, with an energy scale that we denote by $m_\phi$ and therefore only the holon modes with higher energy can be safely integrated out.
In this way no singular terms due to the integration of gapless excitations arise and both dynamics and interactions of the order parameter will be approximately local in space. A similar approach was proposed in [25], but directly for the hole and using different additional approximations. We assume that the high-frequency integrated modes are superconducting modes and this assumption will be proved to be self consistent later.  The result of integration of these high-energy modes in the presence of the gauge field $A$ can be deduced from the Anderson-Higgs mechanism: the gap equation (\ref{horderp}) for each $\alpha$ has a degenerate manifold of solutions for arbitrary phase $\phi^h_\alpha$, therefore the energy should depend only on gradients of $\phi^h_\alpha$. Then by gauge invariance of the holon-gauge system we obtain in the continuum limit (we take $v_F = 1$ and summation over repeated $\mu=0,1,2$ indices is understood henceforth):

\begin{eqnarray}
\label{phasegaugea}
S_{\text{eff}}^{\Delta}(\phi^h_\alpha, A) \sim 
\sum_\alpha \frac{c_\mu
(\Delta^{h}_{\alpha})}{2} \int d^{3}x
\left(\partial_{\mu} \phi^h_\alpha - 2A_{\mu} + 2 \pi n_{\mu\alpha}\right)^2(x)
\end{eqnarray}

where $c_\mu$ are positive functions and the integer-valued vector currents $n_{\mu\alpha}(\vec{x},t)$ allow $\phi^h_\alpha$ to be self-consistently an angle function taking into account the presence of vortices in the phase of the holon-pair field. Summation on $n_{\mu\alpha}$ is understood in the partition function.

The formalism of integer currents can be made precise with a lattice regularization which we use in the following.
One should finally add the action describing the coupling of the low-energy modes of $\Psi$ with $A$ and $\phi^h$ derived from (\ref{holactintcond}) with UV cutoff $m_\phi$.

\section{Propagator of the holon-pair field}

In this section we argue on the basis of a self-consistent argument that the Euclidean correlation function of the phase of the holon-pair field because of its coupling to a gapless gauge field for $T>T_c$ has a purely exponential decay:
\begin{equation}
\label{deltapropagatortwo}
G_{\Delta}(\vec{x},x_0) \equiv \langle e^{i[\phi^h(\vec{x},x_0)-\phi^h(\vec{0},0)]}\rangle \simeq e^{-m_{\phi}\sqrt{v_\phi^2 x_0^2 + \vec{x}^2}}.
\end{equation}

This behaviour was assumed in [24] in a different setting; here we explain the origin of this behaviour in our approach. (We neglect here for simplicity the index $\alpha$, as it turns out that $m_\phi$ is the same for both values of $\alpha=R,L$.)
The relation (\ref{deltapropagatortwo}) defines the holon-pair phase coherence length $\xi_{\phi} \approx  1/m_{\phi}$. $m_{\phi}$ can be thought as the mass of the phase $\phi^h$ of the holon pairs field and $\xi_{\phi}$ as the mean distance between phase vortices.

The important point of Eq. (\ref{deltapropagatortwo}) is that besides its exponential decay for large $|x|$ it does not show any singularity for small $|x|$ (in particular it is a constant at $x=0$) as it would occur for a true massive field which exhibits Ornstein-Zernike decay with power $|x|^{-1}$ in 3D.
It is the above stressed feature that eventually is responsible for the peaks in the spectral weight precursors of SC, yielding an "effective" gap in the antinodal region for the hole, a mark of pseudogap phenomenology.

We proceed as follows: we first calculate the $G_\Delta$ assuming as action the one obtained in section 4, integrating the high-energy modes of spinons and holons and we show that it is self-consistent with the assumed superconducting-like behaviour of the high-energy modes of the holons. Then in section 6 we calculate perturbatively the effect of $\phi^h$ with such propagator on the low-energy holon modes and we show that the derived formula for the holon propagator  if extended to high energy is consistent with the superconducting behaviour previously assumed.

\begin{figure}[ht]
\centering
\includegraphics[width=12pc]{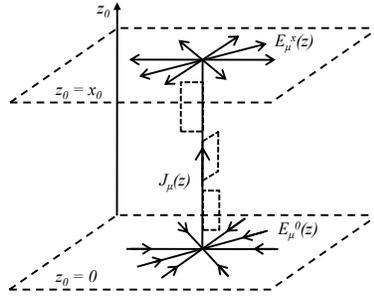}
\caption{The straight temporal current $J_{\mu}(z)$ starting from the point $z=0$, ending at the point $z=x$ and joining the two orthogonal (constant $z_0$) planes which are the support of the electric fields $E_{\mu}^x(z)$ and $E_{\mu}^0(z)$. Suppressed fluctuations of the straight current are also shown.}
\label{fig2EJ}
\end{figure}

To prove Eq. (\ref{deltapropagatortwo}) in the setting described above we evaluate the correlation $G_{\Delta}(x)$  in the Coulomb gauge $\partial_i A_i = 0$ ($i=1,2$ indicates spatial components and $x=(\vec{x},x_0)$) with a lattice regularization. A gauge-fixing is necessary in view of Elitzur theorem [26] that states that without gauge-fixing all non-gauge-invariant correlators vanish.  Due to the presence of the integer vector currents $n$ in (\ref{phasegaugea}) one cannot evaluate it perturbatively. Therefore we use a construction due to Dirac [27] to circumvent the problem: We introduce the gauge invariant field

\begin{eqnarray}
\label{gauinvfield}
e^{i\phi^E(x)} = e^{i[\phi^h(x) + 2 \sum_y E_{\mu}^{x}(y) A^{\mu}(y) ]}
\end{eqnarray}

where $E_{\mu}^x(y)=\delta^1_{x_0}(y_0)(E_1^{\vec{x}}(\vec{y}),E_2^{\vec{x}}(\vec{y}),0)$. Here  $E_i^{\vec{x}}$ is the classical (lattice) 2D electric field generated by a unit charge at $\vec{x}$, hence satisfying the 2D Gauss law  $\partial_i E_i^{\vec{x}}(\vec{y}) = \delta^2_{\vec{x}}(\vec{y})$ and $\delta^d$ denotes the $d$-dimensional Kronecker delta.  Since Eq. (\ref{gauinvfield}) reduces to $e^{i\phi^h(x)}$ in the Coulomb gauge, we can evaluate the correlation $G_{\Delta}(x)$ estimating the expectation value $\langle e^{i[\phi^E(x)-\phi^E(0)]} \rangle$ without gauge fixing, because it is gauge-invariant. Notice that the "electric" field $E_{\mu}^x(y)$ has no temporal component and it is different from zero only in the temporal plane $y_0=x_0$ (see Fig. \ref{fig2EJ}).

Taking into account the periodicity of the phase, i.e. the presence of vortices, we apply the Poisson summation formula to the lattice regularization of (\ref{phasegaugea}) rewriting
\begin{eqnarray}
\label{phasegaugeaca}
\sum_n\exp(-S_{\text{eff}}^{\Delta}(\phi^h,A,n))=\sum_J \exp(-S_{\text{eff}}^{\Delta}(\phi^h,A,J))
\end{eqnarray}
where $J_{\mu}(x)$ are integer vector currents and
\begin{eqnarray}
\label{phasegaugeacal}
S_{\text{eff}}^{\Delta}(\phi^h,A,J)=
\sum_x \left[\frac{J_{\mu}J_{\mu}}{2 c_\mu}+
i J_{\mu}(\partial_{\mu} \phi^h - 2A_{\mu})\right](x).
\end{eqnarray}
Then we have:
\begin{eqnarray}
\label{phiE}
 \langle e^{i [\phi^E(x)-\phi^E(0)]} \rangle=\frac{\int{\cal D}A {\cal D}\phi^h
\sum_J
e^{-S_{\text{eff}}^{\Delta}(\phi^h,A,J)}
e^{i[\phi^E(x)-\phi^E(0)]}}
{\int {\cal D}A{\cal D}\phi^h \sum_J  e^{-S_{\text{eff}}^{\Delta}(\phi^h,A,J)}
},
\end{eqnarray}
where ${\cal D}A{\cal D}\phi^h$ denotes the standard measure for lattice fields.
Defining $J_{x,y}^{\mu}(z)$ a current of charge 2 supported on a path from $x$ to $y$ (i.e. $\partial_{\mu} J_{x,y}^{\mu}(z) = 2(\delta^3_{x}-\delta^3_{y})(z)$), we can write

\begin{equation}
\label{curgammadef}
\phi^h(x)-\phi^h(0) = \sum_z J_{x,0}^{\mu}(z) \partial_{\mu}\phi^h(z)
\end{equation}

and performing the functional integration in $\phi^h$, appearing linearly in the action, we obtain the constraint $\partial_{\mu}J_{\mu}(z) = 2(\delta^3_{x}-\delta^3_{0})(z)$ in the numerator of eq. (\ref{phiE}) and the constraint $\partial_{\mu}J_{\mu}(z) = 0$ in the denominator. The final step is to integrate out the slave particle gauge field $A_{\mu}$. The result at zero temperature is:
\begin{eqnarray}
\label{expgphase}
G_{\Delta}(x) =\frac{
\sum_{ \{J_{\mu}:\partial_{\mu}J_{\mu} = 2(\delta^3_x-\delta^3_0) \} }
e^{-\sum_z \frac{J_{\mu}J_{\mu}}{2 c_\mu}}
e^{-\frac{3 \pi m_s}{2}\sum_{z,w}
[J_{\mu}(z)+ 2 E_{\mu}^{x}(z)-2 E_{\mu}^{0}(z)]\Delta_3^{-1}(z-w)[J_{\mu}(w)+2 E_{\mu}^{x}(w)-2 E_{\mu}^{0}(w)]}
}
{ \sum_{ \{J_{\mu}:\partial_{\mu}J_{\mu} = 0 \} } e^{-\sum_z \frac{J_{\mu}J_{\mu}}{2 c_\mu}} e^{-\frac{3 \pi m_s}{2}\sum_{z,w}
J_{\mu}(z)\Delta_3^{-1}(z-w)J_{\mu}(w)}
}\nonumber \\
\end{eqnarray}
where $\Delta_3^{-1}(z)$ is the inverse of the 3-dimensional lattice Laplacian (at finite $T$ we just replace it by its finite temperature version and restrict the time axis to $[0, 1/T]$ with periodic boundary conditions on the fields). The sum in the numerator is on currents $J_{\mu}$ starting from the point $z=0$ and reaching the point $z=x$ and on closed currents, while only closed currents are present in the denominator. Hence the denominator can be interpreted as the partition function
of a gas of current loops interacting via the 3D (lattice) Coulomb potential. In the numerator the open current has endpoints where the electric current spread out in
fixed time planes, as described by $E$ . The currents $J$  can be interpreted as the
Euclidean worldlines of charge 2 particles. Currents supported on loops correspond
to worldlines of virtual particle-antiparticle pairs, the open current corresponds to the worldline of a particle created at one end of the line
and annihilated at the other one. A similar problem has been dealt with in [28] and the result can be summarized as follows: for sufficiently small $c_\mu$ the closed currents in the leading approximation just weakly renormalize the coefficient of the Maxwell action for $A$, so one can approximate (\ref{expgphase}) setting the denominator to 1 and retaining in the sum of the numerator only a single fluctuating open current $J$ with weighting factor
\begin{eqnarray}
 \label{expgphase1}
e^{-\sum_z \frac{J_{\mu}J_{\mu}}{2 c_\mu}}e^{-\frac{3 \pi m_s}{2}\sum_{z,w}
[J_{\mu}(z)+ 2 E_{\mu}^{x}(z)-2 E_{\mu}^{0}(z)]\Delta_3^{-1}(z-w)[J_{\mu}(w)+2 E_{\mu}^{x}(w)-2 E_{\mu}^{0}(w)]}.
\end{eqnarray}
The factor $e^{-\sum_z \frac{J_{\mu}J_{\mu}}{2 c_\mu}}$ produces an exponential decay in $|x|$. The power decay of the correlation function is then decided by how strongly the current $J$ can fluctuate. Gaussian fluctuations would produce a power of $|x|^{-1}$, however in our case the Coulomb potential in 3 (Euclidean) or less dimensions is confining. Every fluctuation away from a current of minimal length can be described as the addition of a closed loop to the minimal current. Because of confinement such loop produces an exponential factor decaying with its area and thus strongly suppressing every deviation from a current of minimal length. This forces the fluctuations of the current to be non-gaussian, lying within a thin tube surrounding the shortest straight path (see Fig. \ref{fig2EJ}). An approximate evaluation of Eq. (\ref{expgphase1}) is given in the Appendix and it reproduces Eq. (\ref{deltapropagatortwo}) with a $m_{\phi}$ depending on $T$ through $\Delta^h(T)$ (contained in $c_\mu$).
The role of the gauge field $A$ was crucial in obtaining this result. Without it the open current in (\ref{expgphase}) would have gaussian fluctuations and the correlation would exhibit the standard Ornstein-Zernike decay of a free massive field.

\begin{figure}[ht]
\centering
\includegraphics[width=12pc]{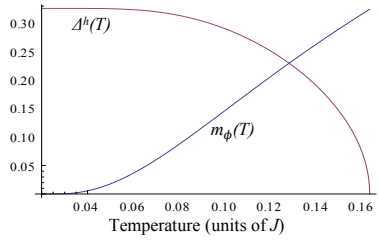}
\caption{$\Delta^{h }(T)$ and $m_{\phi}(T)$  for $\delta=0.18$}
\label{fig3Mph}
\end{figure}

 In Fig. \ref{fig3Mph} we show the typical temperature behaviour of $\Delta^{h }(T)$ and $m_{\phi}(T)$ obtained with the approximations described in Appendix .

\section{Holon self energy and self-consistency}

In this section we extend the discussion of the previous one about holon pairing considering the scattering of holon quasi-particles against the fluctuations of the phase of the holon-pairs field. This section was inspired by [24], however here we give an explicit analytical formula interpolating from the Fermi-liquid and the superconducting behaviour of holons which differs from the one proposed there and in particular our treatment of inclusion of dissipation is completely different from that adopted there.
We exploit the action in Eq. (\ref{holactintcond}) and Eq. (\ref{deltapropagatortwo}) to evaluate, in the continuum limit, the zero temperature self energy $\Sigma (\omega,\vec{k})$ of holon quasi-particles.

What follows holds both in the PG region and in the SM region since the FS shape is irrelevant; for concreteness we discuss the SM case. The difference between the two regions lies in the value of the involved parameters, essentially $k_F$.
The only condition we need is parity invariance which implies that the quasi-particle excitation energy verifies $E(\vec{k})=E(-\vec{k})$.

We consider the holon right + field $\Psi_{+,R}$, the left $\Psi_+$ is similar, with $\gamma_R(\vec{k})\rightarrow\gamma_L(\vec{k})$. Also the $\Psi_-$ field can be treated along similar lines with obvious changes. In the limit $|\Delta^h| << \epsilon_F$ the self energy reads

\begin{equation}
\label{sengone}
\Sigma (\omega,\vec{k})= |\Delta^h|^2 \gamma_R(\vec{k})^2 2\pi m_{\phi}\frac{d}{dm_{\phi}^2} I(k),
\end{equation}

\begin{equation}
\label{idikappadef}
I(k)= \int\frac{d^3q}{(2\pi)^3} \frac{1}{q^2+m_{\phi}^2} \frac{1}{i(q_0-k_0)-E(\vec{q}-\vec{k})}.
\end{equation}
where the $\vec{q}$ dependence of $\gamma_R$ has been neglected for $|\vec{q}|<<|\vec{k}| \sim k_F >> m_{\phi}$ and
 we used $G_\Delta (q_0,\vec{q})=(8 \pi m_{\phi})/(q^2+m_{\phi}^2)^2$.
Considering quasi-particles on a shell of thickness $2\Lambda << k_F$ about the $FS$ we can linearize the quasi-particle dispersion and assuming $\vec{q}^2 << \vec{k}^2$ we have $E(\vec{q}-\vec{k})=v_{F}(|\vec{q}-\vec{k}|-k_F)\approx v_{F}( - \vec{q} \cdot \hat{k} + \delta k)$, where  $\delta k \equiv |\vec{k}|-k_F$, obtaining:

\begin{eqnarray}
\label{inttwo}
I(k) \approx \frac{1}{4\pi}\frac{\sqrt{k_0^2+\delta k^2+m_{\phi}^2}-m_{\phi}}{ik_0+\delta k}
\end{eqnarray}

Inserting Eq. (\ref{inttwo}) into Eq. (\ref{sengone}) and replacing $\delta k$ by $E(\vec{k})$ we obtain for the self energy

\begin{eqnarray}
\label{selfengone}
\Sigma (\omega,\vec{k})=
|\Delta^h|^2 \gamma_R(\vec{k})^2
\frac{1}{i\omega+E(\vec{k})}\left[1-\frac{m_{\phi}}{\sqrt{\omega^2+E(\vec{k})^2+
m_{\phi}^2}} \right]
\end{eqnarray}

where $\gamma_R(\vec{k})^2 \approx 2 k_F^2 \sin^2\theta_{\vec{k}}$ and $\theta_{\vec{k}}$ is the angle between the vector $\vec{k}$ and the nodal direction $\vec{Q}_{1}$ in the right sector. The structure of the self-energy (\ref{selfengone}) is reminiscent of the electron Green functions of one-dimensional models with dynamically generated mass, when the charge and spin velocity coincide [50].

The quasi-particle Green's function of the holon is given by
\begin{eqnarray}
\label{holgreen}
G(\omega,\vec{k})&=& [i\omega-E(\vec{k}) - \Sigma (\omega,\vec{k})]^{-1}
\nonumber\\
&=& \frac{1}{ \left\{1 + |\Delta^h|^2\gamma_R(\vec{k})^2
\frac{1}{\omega^2+E(\vec{k})^2}
\left[1-\frac{m_{\phi}}{\sqrt{\omega^2+E(\vec{k})^2+m_{\phi}^2}}\right] \right\}
[i\omega-E(\vec{k})] }.
\end{eqnarray}
This quite complicated expression is our key result. We notice that at zero frequency, $\omega=0$, the only pole of the Green's function $G(\omega,\vec{k})$ is located at $E(\vec{k})=0$, that is at the FS. This means that there is always a FS except exactly at $m_{\phi}=0$, which occurs only in the superconducting phase. For the existence of the FS the second term in (\ref{selfengone}) is crucial, without it a gap would appear. The absence of the gap is consistent with the fact that in spite of a non-vanishing $|\Delta^h|$ no phase transition has occurred.

To understand the meaning of $G(\omega,\vec{k})$ we analyze the two limits of low ($\omega << m_{\phi}$) and high ($\omega >> m_{\phi}$) frequency, which coincide with those in [24].

For $\omega << m_{\phi}$, we expand the self energy in powers of $\frac{\omega}{m_{\phi}}$ up to the second order and the Green's function becomes

\begin{equation}
\label{gf1}
G(\omega,\vec{k})\simeq\frac{Z_{\Delta}(\vec{k})}{i\omega - E(\vec{k})}, \quad
Z_{\Delta}(\vec{k})=\frac{1}{1+\frac{|\Delta^h|^2}{2 m_{\phi}^2}\gamma_R(\vec{k})^2}.
\end{equation}

    Thus for low frequencies the effect of holon pairing appears through the wave function renormalization  weight $Z_{\Delta}(\vec{k})$. The system behaves like a FL with unchanged Fermi velocity $v_F$ and FS of the Nearly Free Electron approximation, but with a strongly direction dependent weight $Z_{\Delta}(\vec{k})$ that heavily suppresses (if $|\Delta^h| >> m_{\phi}$) quasi-particles in antinodal directions reducing the effective FS.
For $\omega >> m_{\phi}$, we expand the self energy in powers $\frac{m_{\phi}}{\omega}$ and we get the Green's function

\begin{eqnarray}
\label{gf2}
G(\omega,\vec{k})\simeq - \frac{i\omega+E(\vec{k})}{\omega^2 + E(\vec{k})^2 + |\Delta^h|^2\gamma_R(\vec{k})^2}
\end{eqnarray}

    which is the quasi-particle Green's function of a $p$-wave superconductor. Thus for high frequencies the holon system behaves like a $d$-wave superconductor in the MBZ, the $d$-wave being obtained gluing the $p_x-p_y $ and $p_x+p_y$ behaviours in $D_R$ and $D_L$, respectively.

 As soon as $m_{\phi} > 0$, the FS and the FL holon quasi-particle peak of weight $Z_{\Delta}(\vec{k})$ appear for low frequencies and the system behaves like a metal. When $ m_{\phi} \lesssim |\omega|$, quasi-particles begin to scatter strongly with the quanta of the phase of the order parameter changing significantly both quasi-particle dispersion and coherence. FL quasi-particles modes are suppressed in favor of $d$-wave SC modes that become much more coherent and relevant. This means that decreasing the value of $m_{\phi}$ one drives the behaviour of the holon system from metallic towards superconducting.
Therefore the energy scale $m_\phi$ separates FL from SC modes, consistently matching with the cutoff in section 4. In fact, fermionic modes of energy higher than $m_\phi$ were previously assumed to be SC-like and had been integrated out to give dynamics to the holon-pair field, see eq. (\ref{phasegaugea}). In this way one can justify the assumption on high energy modes made in sections 4 and 5 to compute the propagator of the phase field, because the obtained self-energy is consistent with the assumption made.

Performing the analytic continuation $i\omega\rightarrow \omega+i\eta$ in Eq. (\ref{holgreen}) we obtain the retarded Green's function
\begin{equation}
\label{holgreenret}
G^R(\omega,\vec{k})=
\frac{\omega+i\eta + E(\vec{k})}{(\omega+i\eta)^2 - E(\vec{k})^2
- |\Delta^h|^2\gamma_R(\vec{k})^2
\left[1-\frac{m_{\phi}}{\sqrt{E(\vec{k})^2+m_{\phi}^2-(\omega+i\eta)^2}}\right] }
\end{equation}
and its imaginary part is proportional to the spectral weight $A(\omega,\vec{k})= - \frac{1}{\pi} \Im G^R(\omega,\vec{k})$ of the holon.

Since there are FL modes, the scattering rates for holons, $\eta$, is dominated by Reizer singularity and can be computed as in [29]. Combining (\ref{holgreen}) with Ioffe-Larkin rule one can show that the reduction of the spectral wight reducing temperature yields a deviation from below of the $T$-linear behaviour of in-plane resistivity, typical of the SM , as discussed in [30].

Finally we should remember that the continuum fields whose spectral weight has been discussed above should eventually be converted again to the original holon field using (\ref{holpsif}), obtaining in $D$ in this approximation
	 				
\begin{eqnarray}
\label{hpp}
\langle h^* h \rangle (\omega,\vec{k})=\left(\langle \Psi_{+,R}^* \Psi_{+,R} \rangle (\omega,\vec{k})+
 \langle \Psi_{+,L}^* \Psi_{+,L} \rangle (\omega,\vec{k})\right) \chi(\vec{k} \in {\rm MBZ})\nonumber\\+\langle \Psi_-^* \Psi_- \rangle (\omega,\vec{k}) \chi(\vec{k} \notin {\rm MBZ}),
\end{eqnarray}
where $\chi$ denotes the characteristic function.
 As a consequence of (\ref{hpp}) the spectral weight of the holon coming from the $\Psi_+$ fields does not have peaks outside the MBZ, i.e. it only exhibits Fermi arcs along the original FS, except near the MBZ boundary where the FS of the NFE approximation is distorted. The same is true inside the MBZ for the $\Psi_-$ field, whose spectral weight therefore exhibits peaks only outside the MBZ.
This suppression of the spectral weight in the outer "shadow" FS is consistent due to the destruction, produced by the fluctuations of the phase field $\phi^h$, of the long-range order appearing in the BCS approximation for the holon pairing. A similar suppression in the spin-density-wave approach was found in [31], but there was due to fluctuating local anti-ferromagnetic order. In this respect physically more similar to our approach is the phenomenological model of [32], where the suppression of the outer part of the FS is also  due to pairing. However, there mathematically this is realized through the presence of a zero in the electron Green function, as in the cluster DMFT treatment of the 2D Hubbard model [33], which has no direct counterpart in our holon Green function.
As a result, in those approaches the spectral gap in the antinodal region is asymmetric w.r.t. the Fermi energy in contrast with the symmetric behaviour found in our approach below $T_{ph}$ in the SM.

\begin{figure}[ht]
\centering
\includegraphics[width=14pc]{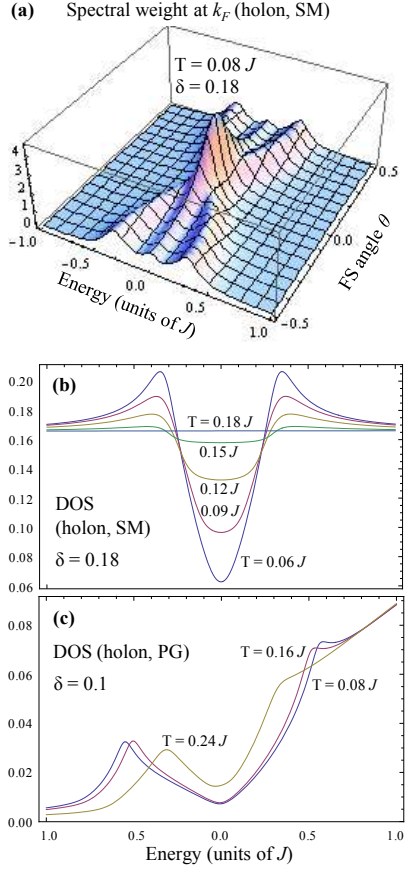}
\caption{Panel (a) shows the holon spectral weight (imaginary part of Eq. (\ref{holgreenret})) at $\vec{k}=\vec{k}_{F}$ versus energy and the FS angle $\theta$, in units of $\pi$, in the SM at fixed $\eta=0.05 J$; the related holon DOS for different values of $T$ is shown in panel (b). Panel (c) shows the holon DOS in the PG for different values of $T$ at the same $\eta$. }
\label{fig4HO}
\end{figure}

Assuming  the Green's function in eq. (\ref{holgreenret}) we plot in Fig. \ref{fig4HO} panel (a) the holon spectral weight at different points of the FS of $\Psi_+$ in the SM with fixed small $\eta$.
The spectral weight is symmetric in frequencies, and exhibits three maxima, the FL-like peak at $\omega=0$ and two symmetric SC-like peaks roughly at

\begin{equation}
\label{omscpemax}
\omega_{scp} \sim \pm \sqrt{|\Delta^h|^2\gamma_R(\vec{k})^2 + m_{\phi}^2}.
\end{equation}

Approximately (al least for small $\eta$) as long as $\frac{\eta}{m_{\phi}} < \frac{m_{\phi}}{|\Delta^h|\gamma_R(\vec{k})}$, the FL peak is higher than the SC peaks which become negligible for $\omega_{scp} >> |\Delta^h|\gamma_R(\vec{k})$, while approximately for $\frac{\eta}{m_{\phi}} > \frac{m_{\phi}}{|\Delta^h|\gamma_R(\vec{k})}$, the SC peaks are higher than the FL one. The size of the area under the peaks determines the main behaviour of the system.

Even if small, the FL peak is always present unless $m_{\phi}$ strictly vanishes and in this case the spectral weight at the FS does not vanish only because of $\eta$. A key point is the direction dependence of the condition on the maxima, due to $\gamma_R(\vec{k})$, which determines the effectiveness of the FS:

\begin{equation}
\label{maxcondspw}
\eta |\Delta^h|\gamma_R(\vec{k}) <[>] m_{\phi}^2,   
\end{equation}
corresponds to a mainly FL[SC] behaviour, respectively.
Indeed near the nodal region the behaviour is always FL-like with an effective FS while near the antinodal directions most of the spectral weight is concentrated on the SC-like modes and even if there is a FS, it has negligible effects.

The density of states (DOS) of holons $DOS_h(\omega)$ can be obtained by numerical integration of the spectral weight $A(\omega,\vec{k})$ in momentum space. Fig. \ref{fig4HO} panel (b) shows the result of this integration in the previously defined momentum shell around the FS for different values of $m_{\phi}$, using the imaginary part of the retarded Green's function in Eq. (\ref{holgreenret}) with fixed $\eta$.

We see that decreasing the value of $m_{\phi}$ the flat FL density of states gradually reduces for small frequencies and develops a peak at $\omega \approx \omega_{scp}$. It is a precursor of the holon SC peak. Higher frequencies are not affected and preserve a flat DOS. As expected, for small $m_{\phi}$ the DOS resembles that of a $d$-wave superconductor with well defined SC-like peaks, but with a finite value at $\omega=0$ due to the FL quasi-particles which, until  $m_{\phi}$ vanishes, preserve a FS, even if not very effective.

As shown in section 7, the behaviour strongly dependent on the value of $m_\phi$ of the holon spectral weight leads to a smoother one for the hole spectral weight, where the hole is obtained coupling the holon to the spinon through gauge fluctuations, because this coupling introduces an additional scattering rate that broadens the peaks appearing in the spectral weight of the holon.
 
A brief comment for the PG region: the holon spectral weight is somewhat similar to that in the SM, but, as shown in Fig. \ref{fig4HO} panel (c), due to the linear dispersion the holon DOS in the PG exhibits a background linearly increasing with $\omega$, besides a dip and two peaks analogous to those in the SM.

We conclude noticing that if we had neglected the Maxwell-like slave-particle gauge field $A_\mu$ in the computation of the correlation function of the holon-pair field, $G_\Delta$ in section 4, the fluctuations of the currents $J_\mu$ defined in section 5 would have been gaussian and we would have obtained the standard massive Ornstein-Zernike behaviour instead of eq. (\ref{deltapropagatortwo}). In this case, using the result in eq.(\ref{inttwo}), we can again evaluate the self-energy and the Green function of the quasi-particles. The resulting density of states is still reduced decreasing $m_\phi$ at low frequencies, however the two superconducting peaks observed above $T_c$ are now nearly absent. As we shall see in section 7, the SC-like peaks for the holons produce analogous peaks in the density of states of physical holes. Their strong reduction appearing for standard fermions, due to the absence of the gauge field, disagrees with experimental data of tunneling in the cuprates. This argument points out, although indirectly, the physical relevance of the slave-particle gauge field.

\section{Reconstructing the hole}

   The Green function $G$ for the hole is given in
coordinate space by the product of the holon and the spinon propagators, averaged over gauge fluctuations.
If we now reinsert the gauge fluctuations by Peierls substitution, since the holon modes with frequency $|\omega| \lesssim m_{\phi}$ basically behave as in a FL, their integration provides a Reizer singularity in the gauge propagator, thought with a reduced coefficient in the effective action w.r.t. the one appearing in absence
of holon pairs. The higher frequency holon modes and the spinons yields only a subleading Maxwell-like correction.
 To obtain a bound state, however, we need to take into account the gauge effect non-perturbatevely. To achieve this goal, of course in approximate form, in [34] we applied a kind of eikonal
resummation of (transverse) gauge fluctuations; details can be found in the above reference, here we just outline the key ideas involved, to explain later on the modifications needed for our case.
The resummation is
obtained by treating first $A_\mu$ as an external field, expanding the correlation function in terms of
first-quantization Feynman paths, then integrating out $A_\mu$ to obtain an interaction between paths
which is then treated in the eikonal approximation. Finally a Fourier transform is performed to
get the retarded correlation function. Whereas the path-representation is straightforward for spinons since they are massive, implementing it on the holons that have a FS causes problems, due to the presence of an arbitrary number of closed fermion wordlines, describing the contributions
of holons in the finite-density ground state. To overcome this difficulty, we apply a dimensional reduction by means of the tomographic
decomposition introduced by Luther and Haldane [35] to the $\Psi$ fields and then gauge it by minimal coupling. To treat
the low-energy holon degrees of freedom we choose a slice of thickness $\Lambda = k_F / \lambda, \lambda>>1$ in
momentum space around the FS of the holon. We decompose the slice in approximately
square sectors; each sector corresponds to a quasi-particle field in the sense of Gallavotti-
Shankar renormalization [36],[37]. Each sector is characterized by a unit vector $\vec{n}(\theta)$, pointing from the center of the FS to the center of the box, labelled by the angle $\theta$ between this direction and the $k_x$ axis. The
contribution of each sector can be viewed approximately as arising from a quasi 1D chiral
fermion; this avoids the finite FS problem for the path-integral representation discussed above.
The paths appearing for a sector are straight lines directed along the Fermi momenta of the
sector, with small Gaussian transverse fluctuations. We apply the eikonal resummation of gauge fluctuations to the composition of the paths for spinons and those arising from holons in each sector. Fortunately the contribution coming from the gaussian fluctuations in a sector turn out to be subleading, so in a sector one can use Gorkov approximation for holons, i.e. the product of the free Green function time a straight phase factor $\exp(i \int_x^y A_\mu dx^\mu)$.
Let us turn to our case. To derive a gauged version of Luther-Haldane decomposition for our holon Green function (\ref{holgreen}) appears too complicate, so we proceeds as follows: we approximate (\ref{holgreen}) with its limiting behaviours:
(\ref{gf1}) for $\omega < m_{\phi}$, (\ref{gf2}) for $\omega > m_{\phi}$. In both cases one can perform the Luther-Haldane decomposition, in the second case one should combine a sector with its opposite to construct a massive quasi 1D fermion [37] and we evaluate the large scale behaviour in approximate form.
For each sector the Gorkov approximation gives again the leading term. The final result of the resummation of the gauge interaction  can then be approximately described for momenta near the FS by four effects modifying the holon Green function (\ref{holgreen}):

1) the most important one is that the hole inherits from the scattering of gauge fluctuations against the spinon, exactly as discussed above in the PG and the SM, a renormalized scattering rate, given  for $T/\omega>>1$ or $<<1$ by
\begin{equation}
\label{Gamma1}
\Gamma(T,\omega)\approx ({\rm Max}(T,\omega) K_F/t) Q(T,\omega)/m_s^2,
\end{equation}
in the SM, where
\begin{equation}
\label{reizer1}
Q(T,\omega)\approx ({\rm Max}(T,\omega) K_F^2)^{1/3},
\end{equation}
is the Reizer momentum,
and
\begin{equation}
\label{Gamma2}
\Gamma(T,\omega)\approx ({\rm Max}(T,\omega) K_F/t)/m_s
\end{equation}
in the PG.
In both cases we use the numerical parameters adopted also in refs. [30] and [34], respectively.
 The broadening due to $\Gamma(T,\omega)$ strongly suppresses the contribution of the FL holon peak away from the diagonal of the BZ.
Here $K_F$  is the (average) Fermi momenta of the FS in absence of holon pairing, because the contributions to the gauge action come from all the pieces of the FS of the holon $h$. In the computation we use an interpolation between the two limiting behaviours in (\ref{reizer1}) and (\ref{Gamma1}),(\ref{Gamma2}) to avoid non-smooth behaviour in the plots.


2) a wave-function renormalization constant $Z(T,\omega) = (K_F m_s Q(T,\omega))^{1/2}$

3) a renormalization of the chemical potential of the holon by adding $m_s$ to it; we denote by $E_\Psi( \vec{k})$ the (renormalized) hole dispersion

4) a sum of two copies of the Green functions obtained as above with  $E_\Psi( \vec{k})$ shifted by $2m_s$.

The effect 3) is due to the fact that the spinon mass $m_s$ naively would produce a gap for the hole. As discussed in the PG [34], since the gauge symmetry is unbroken, the physical hole must have a Fermi surface and this is achieved by a renormalization of the (bare) Fermi momentum which cancel the term $m_s$, guaranteeing a pole at $\omega=0$ in the hole Green function.

The effect 4) is due to the absolute value of $x^0$ in the "relativistic"-like propagator of AF spinon, or equivalently to its double-branch dispersion. This phenomenon is absent in the standard slave-boson theory because the bosonic holon has "non-relativistic" single-branch dispersion. This produces a second term for the hole Green function.
 As a consequence of the shift in the chemical potential discussed above, however the pole in the second Green function is shifted with respect to the Fermi energy by $2 m_s$, which is approximately the mass of the magnon resonance [3].
One can view the second term in the Green function as describing an electron-magnon resonance, due to the "relativistic" nature of the spinon. A phenomenologically similar effect appears in the spin-fermion approach [38].  As explained in a slightly different framework in [39], this kind of particle-hole symmetry breaking is consistent with the original Gutzwiller projection of the $t$-$J$ model. However in our case this asymmetry comes from the coherent contribution and it adds to the one arising from the incoherent contribution not considered above.
Finally we derive the retarded correlation for a $\Psi$ field by taking the complex conjugation for $\omega$ negative.
The resulting retarded correlation near the renormalized FS at an angle $\theta$ is given by
\begin{eqnarray}
\label{ghole}
&&G_\Psi[T, \omega, \theta] =
 Z(T,\omega) \left[
  \frac {\omega + i \Gamma(T, \omega)+ E_\Psi( \vec{k}) }{(\omega + i \Gamma(T, \omega))^2  -E_\Psi(\vec{k})^2 -
      \Delta_h^2(T) \sin(\theta)^2 S(T, \omega)} + \right. \nonumber\\
&& \left. \frac {\omega + E_\Psi(\vec{k})+ 2 m_s+ i \Gamma(T, \omega)  }{(\omega + i \Gamma(T, \omega))^2  - (E_\Psi ( \vec{k})+2 m_s )^2
      -\Delta_h^2(T) \sin(\theta)^2 S'(T, \omega)} \right]
\end{eqnarray}
where
\begin{eqnarray}
\label{S}
S(T, \omega) = 1 - \frac{m_\phi(T)}{(m_\phi(T)^2 +E_\Psi(\vec{k})^2 - (\omega + i \Gamma(T,\omega))^2)^{1/2}}\nonumber\\
S'(T, \omega) =
 1 - \frac{m_\phi(T)}{(m_\phi(T)^2 +(E_\Psi ( \vec{k})+2 m_s )^2 - (\omega + i \Gamma(T,\omega))^2)^{1/2}}.
\end{eqnarray}

\begin{figure}[ht]
\centering
\includegraphics[width=14pc]{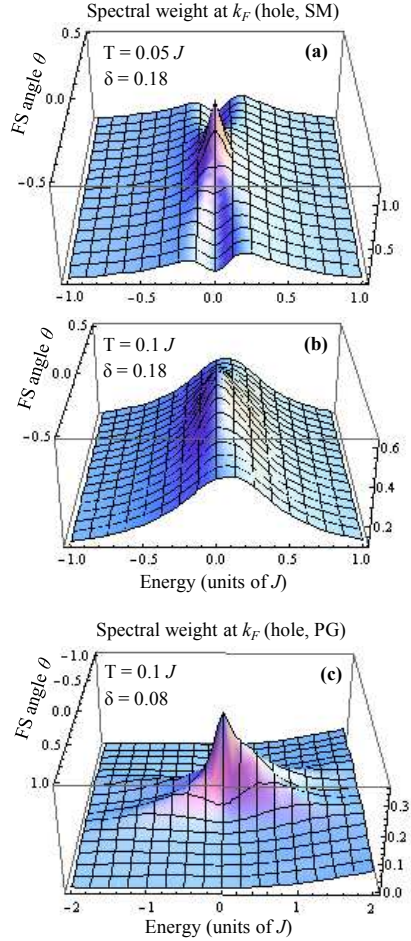}
\caption{Panels (a,b) show the hole spectral weight in the SM (imaginary part of Eq. (\ref{ghole}))and panel (c) in the PG  at $\vec{k}=\vec{k}_{F}$  as a function of the FS angle $\theta$ (in units of $\pi$) and energy. Holes fill the positive energy side.}
\label{fig6WH}
\end{figure}

Features of the resulting spectral weight are shown in Fig. \ref{fig6WH} panels (a) and (b), where the asymmetry and the FL/SC-like crossover with angle and temperature inherited from the holon Green function are evident.

Finally we have to write the full hole retarded Green function combining together using (\ref{hpp}) those originated from the various $\Psi$-fields considered above, obtaining a Green function that exhibits a FS close to the one appearing in the SM without holon pairing, but with strongly modified spectral weight.
This Green functions qualitatively inherits from the holon Green function the pseudogap features discussed in section 6, but with a much stronger scattering rate inherited from the spinon scattering against gauge fluctuations. These pseudogap features are now directly observable in experiments as discussed in section 8, and, contrary to the standard approaches based upon preformed pairs, they appear even if there isn't a gas of preformed hole pairs, since the spin degrees of freedom are not paired.
 The term $S$  due to $m_\phi \neq 0$ is crucial to get a FS for the hole even in the formal limit of vanishing scattering rate $\Gamma$. If one sets  $m_\phi = 0$, hence $S=1$, (and neglect Z and the second term) in (\ref{ghole}) the structure of the Green function is of the kind discussed in [8]. The $\Gamma$ of eqs. (\ref{Gamma1},\ref{Gamma2}) provides the $T$-dependent pair-breaking term introduced phenomenologically there, so many aspects of the pseudogap phenomenology are in common in the two approaches. However the non-trivial $S$ term  due to $m_\phi \neq 0$ is crucial to get a FS without the gap appearing in [8]. This is consistent with the fact that in our approach the appearance of non-vanishing $|\Delta^h|$ occurs from the "normal" states in the PG and SM through a crossover, not a phase transition, which seems consistent with the experimental data.

Let us briefly comment on the situation in the PG. Here there are two factors of suppression for the hole spectral weight. There is the suppression outside the MBZ, due to angle-dependent wave-function renormalization, $Z(\theta)$, consequence of the Dirac structure of holons, and the suppression of the spectral weight away from the diagonals due to holon pairing, which is present in the whole PG region, since $T^* < T_{ph}$. The combination of the two effects leads to a spectral weight concentrated, on the FS, near the nodes of the SC phase. A qualitative difference with respect to the SM is that in the PG there is essentially no "coherent" spectral weight in the antinodal region, since there no FS segments are present.

Finally let us recall that only the "coherent" term due to the hole "resonance" was taken into account in the above discussion. For relative momenta between holon and spinon larger than Reizer's $\sim Q$ and/or energies larger than $\sim v_F Q$ the gauge attraction is unable to bind them and one can treat holon and and spinon approximately as non-interacting. Their contribution to the hole Green function is therefore obtained as a convolution of their free Green functions, producing an incoherent background growing for $\omega >0$ (see section 8).
A typical hole spectral weight in the PG is presented in Fig.\ref{fig6WH} panel (c).

\section{Comparison with experiments}
\begin{figure}[ht]
\centering
\includegraphics[width=14pc]{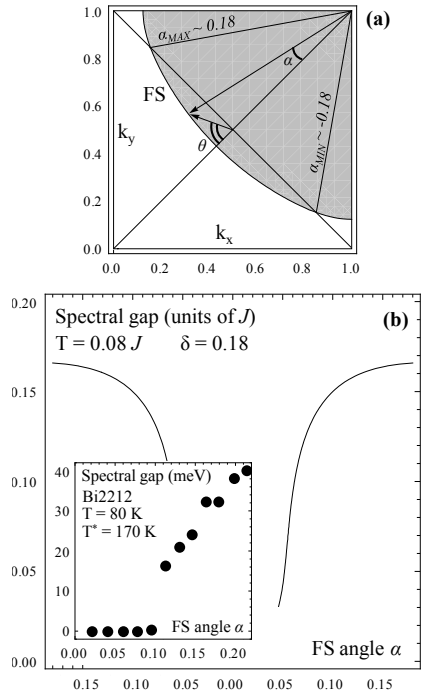}
\caption{Panel (a) shows the definition of the FS angles $\theta$ and $\alpha$ in the first quadrant of the BZ. Panel (b) shows the spectral gap (half the distance between the SC peaks in the symmetrized spectral weight) versus FS angle $\alpha$ measured in units of $\pi$. In the inset data from [47].}
\label{fig8GA}
\end{figure}

Let us now summarize the main results of previous sections useful
to derive a formula for the intensity of ARPES and tunneling.
We have shown that in our approach below a crossover temperature $T_{ph}$ the holons
start to form a gas of holon pairs as a consequence of a long-range attraction
between spin-vortices centered on holons located on opposite Ne\'eel sublattices.
The Green function of the phase of the pairing field, minimally coupled to the
slave-particle gauge field, exhibits a purely exponential decay,
with gap (inverse correlation length) $m_\phi$ decreasing with $T$.
In turn the scattering of the fluctuations of the phase field against holons produces
in the holon Green function, lowering $T$ and moving away from the diagonals of the BZ, a gradual reduction of the spectral weight for $\omega \lesssim m_\phi$
 and simultaneously  for  $ m_\phi \lesssim \omega$ firstly the formation and then an increase
of two peaks of intensity of the spectral weight corresponding to SC-like holon excitations, precursors of
true superconductivity occurring when $m_\phi=0$. Since the holon-pairing distinguishes the two Ne\'el sublattices,
when it is present the holon fields are naturally periodic in the MBZ, the original FS is then distorted, but only near the MBZ boundary for small modulus of the holon-pair field. If one consider the MBZ represented by the two upper quadrants of the BZ, the holon  exhibit two closed hole-like FS centered at the centers of the two quadrants, the corresponding continuum field being denoted by $\Psi_+$, and an  electron-like FS centered at the corners of the region.
For $m_\phi > 0$ the slave-particle
gauge fluctuations exhibit a typical scale, a sort of anomalous
skin momenta, $Q(T,\omega)\approx ({\rm Max}(T,\omega) k_F^2)^{1/3}$ and in a range $|\omega| \lesssim v_F Q_0$ around the holon Fermi energy
the gauge field
couples spinons to holons producing a hole resonance.
The main contribution to the scattering rate $\Gamma(T,\omega)$ of the hole
is due to scattering of the spinon against gauge fluctuations, with $\Gamma(T,\omega)\approx ({\rm Max}(T,\omega))^{4/3}m_s^2$ in the SM and $\approx {\rm Max}(T,\omega)/ m_s$ in the PG.
For momenta in the MBZ contained in first quadrant of the BZ, the derived explicit form of the retarded correlation for the hole resonance can be found in (\ref{ghole}), where $\theta$ is the angle with vertex the center of the quadrant and the angle is measured from the direction corresponding to the node in the MBZ of the modulus of the holon-pair field. However experimental data are usually plotted in terms of the angle, here denoted by $\alpha$,  with vertex the AF wave vector and measured from the direction of the node quoted above (see Fig. \ref{fig8GA} panel (a)). One can get an approximate analytic expression relating $\theta$ to $\alpha$ within the MBZ, by considering the holon FS without pairing as a circle centered at the AF wave vector, with a radius preserving the area. One obtain thus for $\Psi_+$ an average Fermi momentum $k_F \approx 0.1 (1+ 7 \delta)^{1/2}$ and
\begin{eqnarray}
\label{thetalpha}
\sin^2\theta \approx \frac{\sin^2 \alpha}{1+ \frac{\pi}{4(1 + \delta)}-(\frac{\pi}{1 + \delta})^{1/2} \cos \alpha}.
\end{eqnarray}

\subsection{ARPES}

In our approach the intensity for the electron measured in ARPES experiments due to the coherent contribution (of the above discussed hole resonance) for momenta on the FS is proportional to ${\rm Im} G_\Psi[T, -\omega, \theta]n(\omega)$ where  $n(\omega)$ denotes the Fermi function, and the
minus sign in $G_\Psi$ is due to the fact that the ARPES experiments deals with electrons, not with holes as in our approach.
The decrease of the FL peak in the holon spectral weight as we move away from the diagonal and lower the temperature yields  a similar behaviour
for the ARPES intensity, but strongly smoothed out by the scattering of spinons by gauge fluctuations. The particle-hole asymmetry discussed in section 7 and due to the AF structure of spinons smoothed by the thermal broadening appears as a change of the decreasing slope of the spectral weight at large positive $\omega$.
Although the hole spectral weight in our approach is not symmetric in $\omega$, we can easily take the symmetrization of its positive-$\omega$ side. Often one has interpreted as a measure of the  "spectral gap" on the FS half of the distance between the leading peaks of the symmetrized spectral weigh. One assumes zero gap for the case of a single leading peak and these "gapless" portions of the FS are called "Fermi arcs". In Fig. \ref{fig8GA} panel (b) is presented the above defined "gap" as function of the angle $\alpha$, on the portion of the FS relative to $\Psi_+$, defined as in (\ref{thetalpha}) .
Taking the above  definition of arcs we see that in our scheme their length decreases with doping and lowering the temperature, approximately linearly when $|\Delta^h|$ is approximately constant and $m_\phi$ has a $T$-linear behaviour. However in our approach the appearance of the arcs themselves is just an artifact of their definition, the FS is always entirely present, although with reduced spectral weight, up to the SC transition, where the FS becomes gapped except at "d-wave" nodes. Therefore we propose that the "spectral gap" seen in the SM  of cuprates is due to thermal broadening of the SC-like holon peaks, masking the FL peak. The Fermi arcs then correspond to the region of the FS where the FL peak is unmasked.  In some respect ours is the opposite of the proposal in [1], where it was suggested that the pseudogap phenomenon in the SM should be interpreted as due to thermal broadening masking the underlying physical "d-wave" gap.

In the PG the phenomenology is similar, with three main differences: 1) There is a contribution for the FS of the hole corresponding to the holon FS even outside the MBZ due to the Dirac structure of holons. However, since $T_{ph} > T^*$, when we enter in the PG region from the SM the pairing is already active and this implies that the spectral weight is already strongly peaked near the node in the MBZ. 2)  The interval of energy where is effective the attraction between spinon and holon mediated by the slave-particle gauge field is much shorter in the PG than in the SM. This region is in fact triggered by the Reizer momentum that in the PG is approximately $\delta^{2/3}$ times smaller than the one in the SM. This implies that the contribution of the incoherent part (describing the high energy-momentum contribution of the uncoupled spinon-holon) starts at much smaller energy, measured from the Fermi energy, thus appearing in the range of energies considered here for strong underdoping, in contrast with the SM region.
For  low $T$ the leading incoherent contribution to the hole spectral weight on the FS at energy $\omega$ can be approximately estimated as $DOS_h(\omega-m_s)/2m_s$ if $\omega>m_s$ , where $DOS_h$ is the density of states of the holon. Due to the linear dispersion of the holon in the PG, the $DOS_h$ is approximately linear in its argument and the incoherent component of the hole gives a contribution to the spectral weight at positive $\omega$ for the hole, growing approximately linearly in $\omega$ starting from $m_s$ . This particle-hole asymmetric contribution is of the kind discussed in [39]. Tentatively we conjecture that the increase of the symmetrized spectral weight observed in sufficiently underdoped cuprates moving away from the $\omega=0$ peak (see e.g. [40]) have (at least partially) this origin.
 3) Due to the smallness of $k_F$ w.r.t. $\Delta^h$ for strong underdoping (except near $T^*$) one finds that $m_\phi$ is almost 0, hence phenomenologically the spectral weight is similar to that of a d-wave superconductor with a large thermal broadening due to $\Gamma$, somehow agreeing in the PG with the proposal of [1].

\begin{figure}[ht]
\centering
\includegraphics[width=14pc]{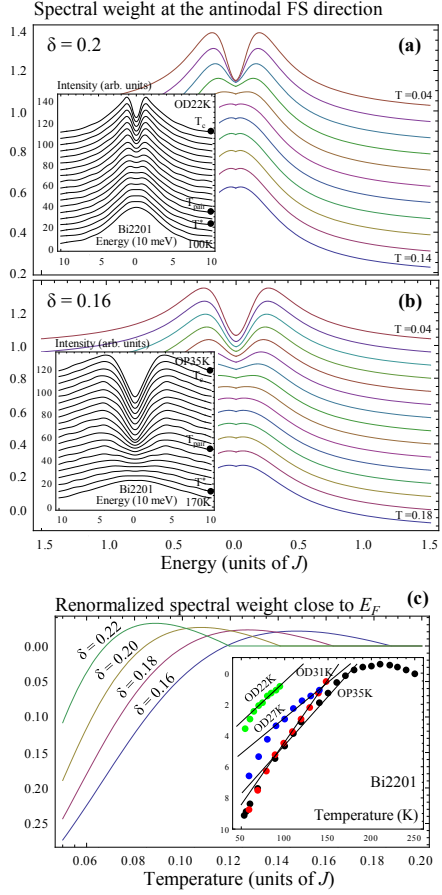}
\caption{Panels (a) and (b) show the symmetrized spectral weight at the "antinodal" FS direction $\theta = 0.45$  in the SM.  The curves are subsequently shifted upward by 0.08 with decreasing $T$ . In the insets experimental data from [41]. Panel (c) shows  the spectral weight at the Fermi energy near the antinode for different doping values, renormalized by subtracting its value at $T_{ph}$. In the inset data from [41]. }
\label{fig9SW}
\end{figure}

\begin{figure}[ht]
\centering
\includegraphics[width=14pc]{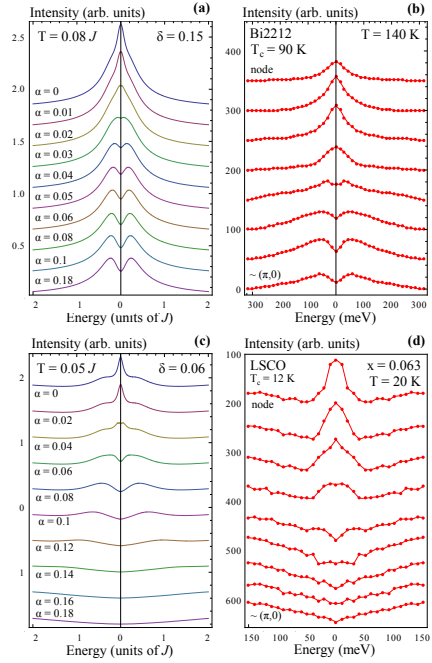}
\caption{Panel (a,c) shows the symmetrized spectral weight as a function of the FS direction $\alpha$ in the SM and in the PG, respectively . Panel (b) shows the experimental data for $Bi2212$ [41] and panel (d) for $La_{2-x}Sr_x CuO_4$ [40]. }
\label{Yoshida1}
\end{figure}

Figures \ref{fig9SW} show the symmetrized spectral weight
 in the SM on the FS at $\theta = 0.45$ for different temperatures. 
The insets present experimental data from [41] exhibiting a good qualitative agreement in terms of angle and temperature dependence with the theoretically derived behaviour. In particular we see in the data that the dip width is reduced moving from underdoped to overdoped samples. We also notice that in underdoped samples, described theoretically by the PG, the position of the SC-like peaks is almost constant with $T$, whereas in optimally-overdoped samples, described by the SM, the peaks are diverging lowering $T$, features well reproduced by our derivation. Furthermore, as shown in panel (c), we found an approximately linear growth with $T$, in some range, of the spectral weight near the antinode on the FS, due to a saddle point in $T$. (In the present treatment doesn't appear the contribution attributed to pair formation in [41] which would arise from the formation of spinon pairs, not considered in the present work.)
The dependence on the FS angle $\alpha$ of the symmetrized spectral weight is pointed out in Fig. \ref{Yoshida1}, panel (a) for SM and (c) for PG and it exhibits a good qualitative agreement with experimental data from [41] for SM (panel (b)) and from [40] for PG (panel (d)).

\subsection{Tunneling}
The (SIN) tunneling intensity is usually proportional to the density of states of the electron. In the cuprates it might be modified by "matrix element" effects, in particular due to orbitals not taken into account by the $t$-$t'$-$J$ model and producing a particle-hole asymmetric term, but close to $\omega = 0$ such perturbing effects are believed not to produce significant qualitative changes [42] and we attempt to compare our theoretical results with experimental data.  To extract some plot to compare with experiments after approximating the holon FS with a circular one with Fermi momentum $k_F$ to simplify the calculation, we integrate the "coherent" part of spectral weight on momenta in a strip of width $\approx Q$ around the FS, to pick up the "coherent" contribution and we have checked numerically that almost all the spectral weight fall in this strip.  For sufficiently small $\omega$  even this very rough approximation reveals some features of DOS that well compare with the data in cuprates.

\begin{figure}[ht]
\centering
\includegraphics[width=14pc]{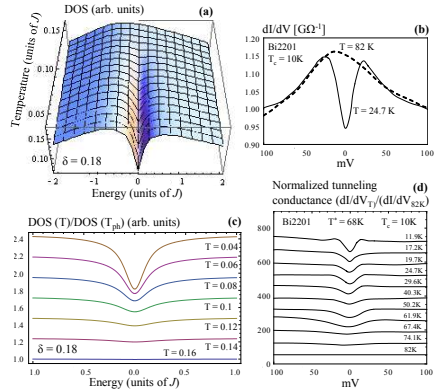}
\caption{(a) The electron DOS in the SM . (c) The same electron DOS renormalized by subtracting its value at $T_{ph}$ at different temperatures.  The curves are subsequently shifted up of 0.04 with decreasing $T$ for clarity. The experimental data (b) and (d) are from [44].}
\label{fig10DO}
\end{figure}

We present in Fig. \ref{fig10DO} (a) the results for the electron DOS obtained lowering the temperature at fixed doping. The DOS exhibits even at high $T$ a broad maximum at $\omega=0$, due to the strong non-  FL dependence on $\omega$ of the scattering rate, furthermore below $T_{ph}$ a dip near $\omega=0$ starts to appear. Both features appear consistent with tunneling experiments [43],[44]  in slightly overdoped cuprates, that should be described by the SM.
In a normalized DOS,  obtained dividing by the DOS at $T_{ph}$, (see fig. \ref{fig10DO} (c) ) the broad maximum disappears.
 Our DOS is qualitatively similar to a phenomenological proposal in [45].
 At fixed temperature as the doping increases the depth of the dip decreases.

A final comment on the PG, here the phenomenology is similar except that the incoherent contribution discussed at the end of previous subsection produces an additional background growing with energy away from $\omega=0$ in the electron side.
 We conjecture that the incoherent contribution is at least partially responsible for the asymmetric behaviour seen in SIN tunneling experiments on sufficiently underdoped samples (see e.g. [46]).

\section{Appendix: Computation of  Eq. (\ref{expgphase1})}
Following the discussion in section 5 we estimate  Eq. (\ref{expgphase1}) in the particular case $x=(\vec{0},x_0)$, retaining only the straight current, and, to simplify the computation, in the continuum limit with ultraviolet cutoff the lattice spacing, $\varepsilon$. The open current has only the temporal component, $J_{\mu}(z) = 2 \delta_{\mu 0} \delta^2(\vec{z})\Theta(x_0-z_0)$, where $\Theta$ is the Heaviside step function, and, because $E_{\mu}$ has no temporal component, we can decouple $J_{\mu}$ from $E_{\mu}$ in the term with the inverse Laplacian in Eq. (\ref{expgphase}) and factorize two contributions to $G_{\Delta}(x)$. The first one is due to the straight current $J_{\mu}(z)$ and the second one is due to the "electric" field $E_{\mu}^{z}$:

\begin{eqnarray}
\label{expgphaseaa}
G_{\Delta}(x_0,\vec{0}) \approx 
e^{-\left[ \frac{1}{2 c_0} x_0 + 6 \pi m_s
\int_0^{x_0}dt\int_0^{x_0}dt'\Delta_3^{-1}(t-t',\vec{0})\right]}e^{-\frac{3 \pi m_s}{2} D(x_0)}
\end{eqnarray}

where
\begin{eqnarray}
\label{intphasee}
D(x_0) = \int d^3z d^3w
[2 E_{\mu}^{x}-2 E_{\mu}^{0}](z)\Delta_3^{-1}(z-w)[2 E_{\mu}^{x}-2 E_{\mu}^{0}](w)=
\nonumber \\
\frac{1}{ 2 \pi^3} \int d^2x d^2y \frac{\vec{x} \cdot \vec{y}}{|\vec{x}|^2 |\vec{y}|^2}
\left[\frac{1}{\sqrt{(\vec{x}-\vec{y})^2+\varepsilon^2}}-
\frac{1}{\sqrt{|\vec{x}-\vec{y}|^2+x_0^2+\varepsilon^2}}\right]
\sim C_1 x_0  + C'\ln (x_0).
\end{eqnarray}
Eq. (\ref{intphasee}) follows from the continuum limit definitions of $E_{\mu}^x(z) = \delta(z_0-x_0) \frac{1}{|\vec{z}-\vec{x}|}(z_1-x_1,z_2-x_2,0)$ and of the inverse 3-D Laplacian $\Delta_3^{-1}(z) = \frac{1}{\sqrt{z_1^2+z_2^2+z_0^2}}$ with $\varepsilon$ as UV cutoff. The last step holds in the limit $x_0 >> 1 >> \varepsilon$ and $C_1$,$C'$ positive constants.

The important point is that the logarithmic contribution in Eq. (\ref{intphasee}) is exactly eliminated from the integral in the first exponential of Eq. (\ref{expgphaseaa}) due to the straight current,

\begin{eqnarray}
\label{expgphasecc}
\int_0^{x_0}dt\int_0^{x_0}dt'\Delta_3^{-1}(t-t',\vec{0})= 
\int_0^{x_0}dt\int_0^{x_0}dt'\frac{1}{\sqrt{(t-t')^2+\varepsilon^2}}
\sim C_2 x_0 - C'\ln (x_0)
\end{eqnarray}
with $C_2$ a positive constant and numerically $C_3 \equiv C_1 + C_2 \approx 25$.
In conclusion $G_{\Delta}(x)$ turns out to  have a purely exponential decay in $|x|$ and from the above discussion, our estimate of the inverse coherence length for small $\Delta^{h }$ is $m_{\phi} \approx \frac{1}{2 c_0} + \frac{3 C_3 \pi}{2} m_s$.

 Actually with the more sophisticated methods of the excitation expansions of [48], one obtains for small $\Delta^{h }$ the estimate

\begin{equation}
\label{stimamass}
m_{\phi} \approx \frac{1}{2 c_0} + \frac{3 C_3 \pi}{2} m_s - P(c_\mu)
\end{equation}

where $P$ is a polynomial taking into account the fluctuations of the current $J_{\mu}(x)$; in an extrapolation at lower temperatures it ensures the right behaviour when the SC temperature for the holon system, $T_{ch}$, is approached and $m_{\phi}$ tends to zero. It should be stressed that $T_{ch}$ is not the physical SC temperature $T_c$, because in the physical system also the spinons contribute to the SC transition through the formation of RVB pairs, so that $T_{ch} \neq T_c$.

In BCS approximation $c_1=c_2\equiv c$ is proportional to the superfluid density, $\sim (\Delta^h)^2$ for small $\Delta^h$ , plus a correction $\sim m_\phi^2$ and $c_0$ is approximately proportional to $k_F/v_\phi$, therefore for sufficiently small $\Delta^h$ and $k_F$ the above computations are justified.
However, to compute using the above procedure the explicit dependence of $m_\phi$ on $T$ needed in the following would be a formidable task. We expect and we have partially verified numerically that the physical results depend only weakly on the explicit form of this dependence and in the computations we adopt the following drastic simplifications: First we replace $v_\phi$ by $v_F$, this would reproduce the correct order of magnitude for an s-wave pairing and according to the results in [49] it is not unreasonable even for a d-wave pairing; anyway we have verified that numerically the precise value is essentially irrelevant in the range of temperatures considered here. Second we keep only the linear term of $P$, with a coefficient chosen exactly to ensure the vanishing of $m_{\phi}$ at the (unphysical) holon superconducting temperature  $T_{ch}$ , assuming that all the temperature dependence is appearing in  $\Delta^{h }(T)$.
We thus derive the following expression for $m_\phi(T)$ used in the plots:
\begin{equation}
\label{explicit}
m_\phi( T) = (-\Delta^h (T_{ch})^2 + (\Delta^h (T_{ch})^4 + 
       4 A^2 (\Delta^h (T_{ch})^2 - \Delta^h (T)^2))^{1/2})/(2 A).
\end{equation}
In (\ref{explicit}) $A =(2 k_F)^{-1} + (75 \pi /2 m_s)$ and (away from $T=0$) we fit numerically $\Delta^{h }(T)$ by the approximate expression $\Delta^{h }(T)/\Delta^{h }(0) \approx [1 - e^{2.5 (1 - T_{ph}/T))}]^{1/2}$ and [2]
$\Delta^h(0)\approx 1.6 Z(\delta) k_F^{1/2} e^{- \frac{0.34}{Z(\delta)} }$
 with $Z(\delta) \equiv (1 - 2 \delta) ((1 + m_s^2)^{1/2} - m_s)$.
 
 Furthermore since the holon system is not in the weak BCS region,  $T_{ph} >> T_{ch}$ and we just set  $T_{ch} \approx 0$.

{\bf Acknowledgments} Useful discussions with J. C. Campuzano, A. Kaminsky, A. Sacuto, L. Salasnich, Z.B. Su, F. Toigo, A. Tsvelik and in particular with L. Yu and F. Ye are gratefully acknowledged.



\end{document}